\documentclass[
  journal=pasa,
  manuscript=article-type,
  year=2025,
  volume=1,
]{cup-journal}

\usepackage{amsmath}
\usepackage[nopatch]{microtype}
\usepackage{bbm}
\usepackage{booktabs}
\usepackage{amssymb}
\usepackage{commath}
\usepackage{graphicx}
\usepackage{subcaption}

\usepackage{tikz,xcolor,hyperref}

\newcommand{\direct}{\textsc{DIReCT}}
\newcommand{\gendirect}{\textsc{GenDIReCT}}

\newcommand{\ehtim}{\textsc{eht-imaging}}

\newcommand{\hops}{\textsc{EHT-HOPS}}
\newcommand{\mnist}{\textsc{MNIST}}
\newcommand{\imagenet}{\textsc{ImageNet}}

\defcitealias{Chael_2018_ehtim}{C18}
\defcitealias{Roelofs2023_ngeht-challenge}{R23}
\defcitealias{Kim_2020_3C279}{K20}
\defcitealias{Janssen_2021_CenA}{J21}
\defcitealias{Lai_2025_GenDIReCT}{L25}

\definecolor{lime}{HTML}{A6CE39}
\DeclareRobustCommand{\orcidicon}{%
    \begin{tikzpicture}
    \draw[lime, fill=lime] (0,0) 
    circle [radius=0.16] 
    node[white] {{\fontfamily{qag}\selectfont \tiny ID}};
    \draw[white, fill=white] (-0.0625,0.095) 
    circle [radius=0.007];
    \end{tikzpicture}
    \hspace{-2mm}
}

\newcommand{\orcidSamuel}{\href{https://orcid.org/0000-0001-9372-4611}{\orcidicon}}
\newcommand{\orcidNT}{\href{https://orcid.org/0000-0003-1602-7868}{\orcidicon}}

\title[\gendirect\ Imaging of CenA and 3C~279]{Conditional Image Diffusion with Interferometric Closure Invariants: Independent EHT Imaging of Centaurus~A and 3C~279}

\author{Samuel Lai\orcidSamuel}
\affiliation{Space \& Astronomy, Commonwealth Scientific and Industrial Research Organisation (CSIRO), P. O. Box 1130, Bentley, WA 6102, Australia}
\email[Samuel Lai]{samuel.lai@csiro.au}

\author{Nithyanandan Thyagarajan\orcidNT}
\affiliation{Space \& Astronomy, Commonwealth Scientific and Industrial Research Organisation (CSIRO), P. O. Box 1130, Bentley, WA 6102, Australia}

\author{O. Ivy Wong}
\affiliation{Space \& Astronomy, Commonwealth Scientific and Industrial Research Organisation (CSIRO), P. O. Box 1130, Bentley, WA 6102, Australia}
\alsoaffiliation{International Centre for Radio Astronomy Research, The University of Western Australia, Crawley, WA 6009, Australia}

\author{Foivos Diakogiannis}
\affiliation{Data 61, Commonwealth Scientific and Industrial Research Organisation (CSIRO), Kensington, WA 6151, Australia}


\begin{document}

\begin{abstract}
We present independent imaging analyses of Event Horizon Telescope (EHT) observations of the active galactic nuclei in radio galaxy Centaurus~A and quasar 3C~279 using Generative Deep learning Image Reconstruction with Closure Terms (\gendirect), a recently developed machine-learning framework built on conditional diffusion models that uses interferometric closure invariants as primary observables. 
For Centaurus~A, our reconstruction reveals two prominent emission ridges ($\simeq 80\,\mu$as each) along the jet sheath with a brightness ratio of $1.4\pm 0.1$ and an opening angle of $12.3\pm 0.3$~deg. For 3C~279, we identify three distinct components in the image, with the southern jet ejecta on sub-parsec scale exhibiting a proper motion of $4.6\pm 1.0\,\mu$as over $\approx 5.39$ days away from the northern components, corresponding to an apparent superluminal velocity of $\simeq 10\pm 2$ times light speed. These measurements are consistent with those reported by the EHT Collaboration. The results are significant because we demonstrate that: (1) imaging from interferometric aperture synthesis data, especially in VLBI and most acutely in extremely sparse arrays like the EHT, remains a severely ill-posed and challenging inverse problem, yet closure invariants preserve robust morphological information that can strongly constrain structural features, and (2) more importantly,  closure-invariant imaging largely avoids calibration systematics, thus providing a fundamentally independent view of spatial structure with very high angular resolution. The generative nature of \gendirect\ further allows us to sample and characterise clusters of plausible image solutions for each dataset. As a calibration-independent, generative imaging approach, \gendirect\ offers a robust and truly independent blind-imaging tool for current and future VLBI experiments.
\end{abstract}

\section{Introduction} \label{sec:Introduction}

Very-long baseline radio interferometry is responsible for achieving the highest angular resolution imaging results in the astronomy discipline. By measuring signal correlations, known as visibilities, from pairs of radio receivers sampling the aperture on continental or global scales, one can infer the source intensity distribution by, in the most straightforward case, a two-dimensional inverse Fourier transform of the measured visibilities \citep{TMS}. However, correlations measured across large spatial distances from a limited array of heterogeneous antennas naturally results in sparse coverage on the aperture plane, which is further complicated by instrumental noise and station-based calibration errors, encompassing amplitude attenuations and phase delays in the signal path both from external influences such as the propagation medium  (troposphere, ionosphere, etc.) and from the electronics in the receiver systems.

Accurate image recovery necessitates precise and reliable restoration of the radiation's coherence via signal  calibration. Yet, the calibration process itself is often a meticulously fine-tuned iterative process that involves a multi-stage optimisation of both the model image and station-based gains \citep{Pearson_1984_SelfCal}. Different assumptions adopted during the calibration and deconvolution process can influence the resulting reconstruction, leading to divergent results and a mixture of scientific interpretations. For instance, in the recent observations of the ring-like structure of M87 \citep{EHT_2019_Data} produced by the Event Horizon Telescope Collaboration \citep[EHTC;][]{Doeleman_2009}, the independent analysis of \citet{Carilli_2022} showed non-negligible variation in the imaged structure produced by a hybrid mapping reconstruction algorithm for different initial models. Likewise, through a markedly different choice of deconvolution window, \citet{Miyoshi_2022_M87} demonstrated an alternative reconstruction without a clear ring that is reportedly similarly consistent with the measured visibility data. Multiple other re-analyses of the EHTC data on M87 \citep[e.g.][]{Arras_2022, Broderick_2022, Lockhart_2022, Muller_2024_closureTraces, Feng_2024}, and even independent imaging teams within the EHTC \citep{EHT_2019_Imaging}, generally produce consistent ring-like morphologies, but finer characteristics (i.e., ring thickness, surface brightness profile, hot spot location, etc.) can vary substantially between different methodologies or prior assumptions, which limits the scientific interpretability of features in the image reconstruction. 

For these reasons, techniques capable of direct imaging from calibration-independent physical observables, free of external fine-tuned hyperparameters, are highly valuable. For decades, it's been known that much of the intricacies of the calibration process can be bypassed by considering specific combinations of interferometric measurements, called ``closure quantities''. Traditionally, closure quantities were constructed in closed triangular loops of stations, which cancels out station-based phase corruption \citep{Jennison_1958,Thyagarajan_2022_CPhase}, and quadrilateral loops to cancel out amplitude corruptions \citep{Twiss_1960}. The closure amplitudes and phases are immune to arbitrarily large station-based multiplicative corruptions, carrying robust information on source properties limited only by additive thermal noise and non-station-based errors \citep{Blackburn_2020, Lockhart_2022}. 
An extension of closure quantities to polarimetric measurements, namely ``closure traces'', was provided in \citet{Broderick_2020}. In further advancement, a more accurate and unified theory of ``closure invariants'' that treats all of them homogeneously is available for co-polar and polarimetric interferometry in \citet{Thyagarajan_2022_CI} and \citet{Samuel_2022}, respectively.

Despite their many desirable characteristics, relying on closure quantities alone instead of traditional visibilities has a few limitations. First, the complete set of closure quantities is always smaller than the full set of visibilities, so some information is inevitably lost, including the absolute amplitude scale and any global phase (overall translation in the image). Second, reconstructing an image directly from closure quantities is an ill-posed inverse problem, because closure quantities relate to the underlying visibilities in a non-linear way.
Regularised maximum-likelihood methods \citep[RML; e.g.][]{Ikeda_2016, Akiyama_2017_polarimetric, Akiyama_2017_imaging, Chael_2018_ehtim, Blackburn_2020} explored an approach where selected regularisation terms provided the essential additional constraining power to allow images to be produced directly from closure quantities \citep[e.g.][]{Chael_2018_ehtim}. Since then, closure quantities have been used as a constraint in multiple variations of regularised maximum likelihood methods, including techniques based on compressive sensing principles \citep[e.g.][]{Mertens_2015, Mueller_2022_doghit, Muller_2024_closureTraces} and genetic algorithms for multiobjective optimisation \citep[e.g.][]{Muller_2023_MOEAD, Mus_2024_PSO}. Despite this, the resulting image reconstructions are still sensitive to the initialisation of the optimisation problem and hyperparameters introduced by the prior, or selection and relative weighting of regularisation terms \citep[e.g.][]{Chael_2018_ehtim, Carilli_2022}, leading to a potentially diverse array of solutions depending on prior assumptions.

Recent advancements in machine learning have demonstrated remarkable performance and fidelity on inverse problems, such as image or video denoising and super-resolution \citep[e.g.][]{Zhang_2017, Rombach_2021_StableDiff, Donike_2025}, attracting widespread adoption in astronomy across multiple domains \citep[e.g.][]{Longo_2019, Huertas_2023}. In the radio interferometric imaging problem, a large variety of machine learning methods have already been employed, from convolutional neural networks \citep[e.g.][]{Sureau_2020, Nammour_2022, Schmidt_2022, Chiche_2023, Terris_2023}, adversarial networks \citep[e.g.][]{Geyer_2023, Rustige_2023}, and normalising flow \citep[e.g.][]{Sun_2022_adeep, Feng_2024}, to denoising diffusion probabilistic models \citep[e.g.][]{Drozdova_2024, Feng_2024, Lai_2025_GenDIReCT}. Recently, \citet{Lai_2025_GenDIReCT} presented a generative diffusion imaging model capable of producing total intensity image reconstructions directly from co-polar closure invariants \citep{Thyagarajan_2022_CI}, 
which is a homogenous and unified alternative to
the traditional closure quantities, carrying identical calibration-independent information. Compared to regularised maximum likelihood, the network of \citet{Lai_2025_GenDIReCT}, called \gendirect, learns the effective prior and regularisation from its training dataset, containing a variety of natural and non-astronomy images, which eliminates fine-tuned hyperparameters from the imaging sequence and the calibration process altogether, serving as a user-input agnostic imaging technique. 

In this work, we present the first application of the \citet{Lai_2025_GenDIReCT} methodology on real EHT observations. The chosen targets are the active galactic nuclei 3C~279 \citep{Kim_2020_3C279} and Centaurus~A \citep[Cen A;][]{Janssen_2021_CenA}, which were observed in 2017 alongside M87. However, despite the public accessibility of the datasets, the reconstructed images from the EHTC have hitherto not been independently verified, to our knowledge, by external groups with alternative imaging techniques. Throughout this work, we present independent reconstructions of both targets and quantitatively compare the results to reference reconstructions from the EHTC, as part of a staged demonstration of the capabilities and reliability of \gendirect\ on real data prior to tackling imaging problems on the event-horizon scale. 

The content of this paper is organised as follows: in Sections \ref{sec:EHT}, we briefly describe the Event Horizon Telescope and its observations of blazar 3C~279 and radio galaxy Centaurus~A. In Section \ref{sec:imaging}, we describe \gendirect, the novel generative deep learning approach to imaging reconstruction with closure invariants. We also describe the validation strategy for trained \gendirect\ models. In Sections \ref{sec:results} and \ref{sec:discussion}, we present the first image reconstructions from \gendirect\ on real EHT observations of 3C~279 and Centaurus~A, comparing the results to reference image reconstructions published by the EHTC. We present a summary of this work and conclusions in Section \ref{sec:conclusion}.  For this study, we adopt a flat $\Lambda$CDM cosmology with H$_{0} = 67.7$~km~s$^{-1}$~Mpc$^{-1}$ and $\left(\Omega_{\rm m}, \Omega_{\Lambda}\right) = \left(0.307, 0.693\right)$ \citep{Planck_2016_cosmo}.

\section{The Event Horizon Telescope} \label{sec:EHT}
The Event Horizon Telescope is a global network of heterogeneous millimeter and sub-millimeter wavelength telescopes separated by long baselines with lengths up to those comparable to the Earth's diameter. The angular scales, $\theta \sim \lambda/D$, probed by the EHT at the nominal operating wavelength of $\lambda \sim 1.3$ mm is $\theta\sim 25\,\,\mu$as, which is capable of spatially resolving the accretion and jet formation mechanisms near the critical boundaries of several nearby supermassive black holes \citep[e.g.][]{EHT_2019_Imaging, EHT_2022_SgrAImaging}.

In 2017, the EHT array was comprised of 8 separate facilities \citep{EHT_2019_Array}: the Atacama Large Millimeter/submillimeter Array \citep[ALMA;][]{Wootten_2009_ALMA, Goddi_2019_ALMA}, the Atacama Pathfinder Experiment telescope \citep[APEX;][]{Gusten_2006_APEX}, the Large Millimeter Telescope \citep[LMT;][]{Hughes_2010_LMT}, the Pico Veleta IRAM 30 m telescope \citep[PV;][]{Greve_1995_PV}, the Submillimeter Telescope Observatory \citep[SMT;][]{Baars_1999_SMT}, the James Clerk Maxwell Telescope (JCMT), the Submillimeter Array \citep[SMA;][]{Ho_2004_SMA}, and the South Pole Telescope \citep[SPT;][]{Carlstrom_2011_SPT, Kim_2018_SPT}. 
As the Earth rotates, the spatial frequencies sampled by the EHT form tracks in the Fourier plane, resulting in a sparsely sampled Fourier transform of the sky with non-uniform sensitivity. 

Despite the challenges, the Event Horizon Telescope Collaboration (EHTC) recently achieved noteworthy scientific results in imaging the black hole shadows of the central supermassive black holes within M87 \citep{EHT_2019_Imaging} and Sgr A* \citep{EHT_2022_SgrAImaging}. The images were obtained following a detailed verification process to mitigate both algorithmic and human biases \citep[e.g.][]{Shepherd_2011_DIFMAP, Akiyama_2017_imaging, Chael_2018_ehtim}. However, the fine details of the reconstruction depend on different assumptions, initial models, and calibration solutions \citep[e.g.][]{Arras_2022, Broderick_2022, Carilli_2022, Lockhart_2022, Muller_2024_closureTraces, Feng_2024}, which underscores the value of independent imaging techniques in strengthening the reliability of VLBI imaging results. 

Outside of the event-horizon scale, the EHT has also been used to observe other active galactic nuclei of interest at 230 GHz, including blazar 3C~279 \citep{Kim_2020_3C279} 
and radio galaxy Centaurus~A \citep{Janssen_2021_CenA}. Notably, although the data is public, the image reconstructions from the EHTC on these targets have, to our knowledge, yet to be independently verified by external groups. In this study, we apply our independent imaging algorithm to both the 3C~279 and Centaurus~A EHT datasets and compare the results to the reconstructions from the EHTC. All of the EHT calibrated data products used in this study are from the publicly available data archive on the collaboration's website\footnote{\href{https://eventhorizontelescope.org/for-astronomers/data}{https://eventhorizontelescope.org/for-astronomers/data}}.

\subsection{Blazar 3C~279}
3C~279 \citep[$z=0.5362\pm0.0004$;][]{Marziani_1996_3c279z} is an archetypal blazar, exhibiting rapid time-variability in structure \citep[e.g.][]{Lister_2018} and flux across a wide range of frequencies \citep[e.g.][]{Larionov_2020_3C279}. It is also one of the first objects with evidence of apparent superluminal motions in the relativistic AGN ejecta revealed by VLBI techniques \citep[e.g.][]{Whitney_1971_superluminal}. Due to its high brightness and the high signal-to-noise ratio (SNR), 3C~279 is an ideal source for robust VLBI fringe detection.

In 2017, observations of 3C~279 were taken with the EHT, interleaved with observations of M87, to independently validate the calibration solution of the M87 image \citep{EHT_2019_Imaging}. The data was taken over four nights (5, 6, 10, 11) in April at two 2-GHz bands centered at 227.1 (low) and 229.1 GHz (high), where unlike with M87, the SPT was able to participate, albeit at high airmass \citep{EHT_2019_Imaging, Kim_2020_3C279}. Additional details of the observation are described in \citet{Kim_2020_3C279}, hereafter referred to as \citetalias{Kim_2020_3C279}. The frequency-averaged and network-calibrated visibility data produced by the \hops\ pipeline \citep{Blackburn_2019_EHTHOPS} is publicly available on the collaboration's archive. 

In \citetalias{Kim_2020_3C279}, the reference image, reflecting the most robust features detected by several independent imaging methods across teams within the EHTC, consists of two distinct bright emission regions, separated by $\sim100\,\mu$as (refer to leftmost panel of Figure \ref{fig:3C279CenA_compare}). The northern structure is extended along the NW-SE direction, perpendicular to the orientation of the large-scale jet, while the relative location and elongation of the secondary southern structure is roughly consistent with the jet direction towards the SW \citepalias{Kim_2020_3C279}. We later refer to the northern structure as the assumed ``core'' and the secondary structure as the ``ejecta'', following the \citetalias{Kim_2020_3C279} interpretation. Dramatic inter-day closure phase variations provide strong evidence for rapid variability in the jet structure and surface brightness distribution. Moreover, inter-day difference images show prominent brightness temperature variations in the nuclear region and significant proper motion in the ejected feature. In \citetalias{Kim_2020_3C279}, the imaging problem for 3C~279 was described as more challenging than M87, as a consequence of the source's extended structure and the relative paucity of intermediate baselines in the EHT array \citep{EHT_2019_Imaging}. Across all of the imaging pipelines, the 3C~279 image was produced within a limited $\sim100\times100\,\mu{\rm as}^2$ field-of-view, while a single large-scale Gaussian was used to capture extended emission \citepalias{Kim_2020_3C279}. More recently, observations of 3C~279 by the space VLBI mission, \textit{RadioAstron}, at 22 GHz were published \citep{Fuentes_2023_RadioAstron}. With its significantly longer (near 10$\times$) space-scale baselines, it achieved angular resolution comparable to that of the EHT and revealed filamentary emission on large ($\sim500\,\mu$as) scales, possibly threaded by helical magnetic field structure. 

\subsection{Centaurus~A}
At a distance of $3.8\pm0.1$ Mpc from Earth \citep{Harris_2010_CenADistance}, Centaurus~A is the closest radio galaxy, hosting a supermassive black hole which drives its exceptionally bright radio emission. The EHT observed Centaurus~A over a six-hour duration track on 10 April 2017 with both low and high bands \citep{Janssen_2021_CenA}. Details of the data processing and imaging procedure are described in \citet{Janssen_2021_CenA}, hereafter referred to as \citetalias{Janssen_2021_CenA}. In brief, the observational data was reduced via two pipelines: \textsc{rPICARD} \citep{Janssen_2019_rpicard} and \hops\ \citep{Blackburn_2019_EHTHOPS}. For this study, we obtain the \hops-reduced data from the EHT collaboration data archive. Imaging in \citetalias{Janssen_2021_CenA} proceeded with a blind imaging challenge undertaken by several individual groups within the EHTC who independently produced twelve images, six of which passed the data fidelity threshold. The six images sourced from a variety of imaging methods converged to the same robust source structure. Subsequently, the \ehtim\ script \citep{Chael_2018_ehtim}, used to image M87\footnote{\href{https://github.com/eventhorizontelescope/2019-D01-02}{https://github.com/eventhorizontelescope/2019-D01-02}}, was applied to further refine the image.

The final image model presented in \citetalias{Janssen_2021_CenA} appears as a narrow, collimated, and edge-brightened jet with an approaching side that extends towards the NE direction and fainter counter-jet towards the SW. A brightness asymmetry is evident with a flux ratio of $R_{\rm{s/n}} = 1.6 \pm 0.5$ between the southern and northern ridgelines, most likely caused by relativistic boosting \citepalias{Janssen_2021_CenA}. With the high dynamic range of the image reconstruction, \citetalias{Janssen_2021_CenA} empricially located the jet apex and measured the jet collimation profile. Furthermore, by testing modified versions of the image model with and without secondary features using the same data processing and imaging pipeline, \citetalias{Janssen_2021_CenA} confirmed that the counter-jet and extended emission features (out to $\sim200\,\mu{\rm{as}}$ from the apex) did not spuriously appear in image reconstructions of simulated observations, lending enhanced confidence in the detection of these fainter features. 

\begin{figure*}
	\includegraphics[width=0.95\textwidth]{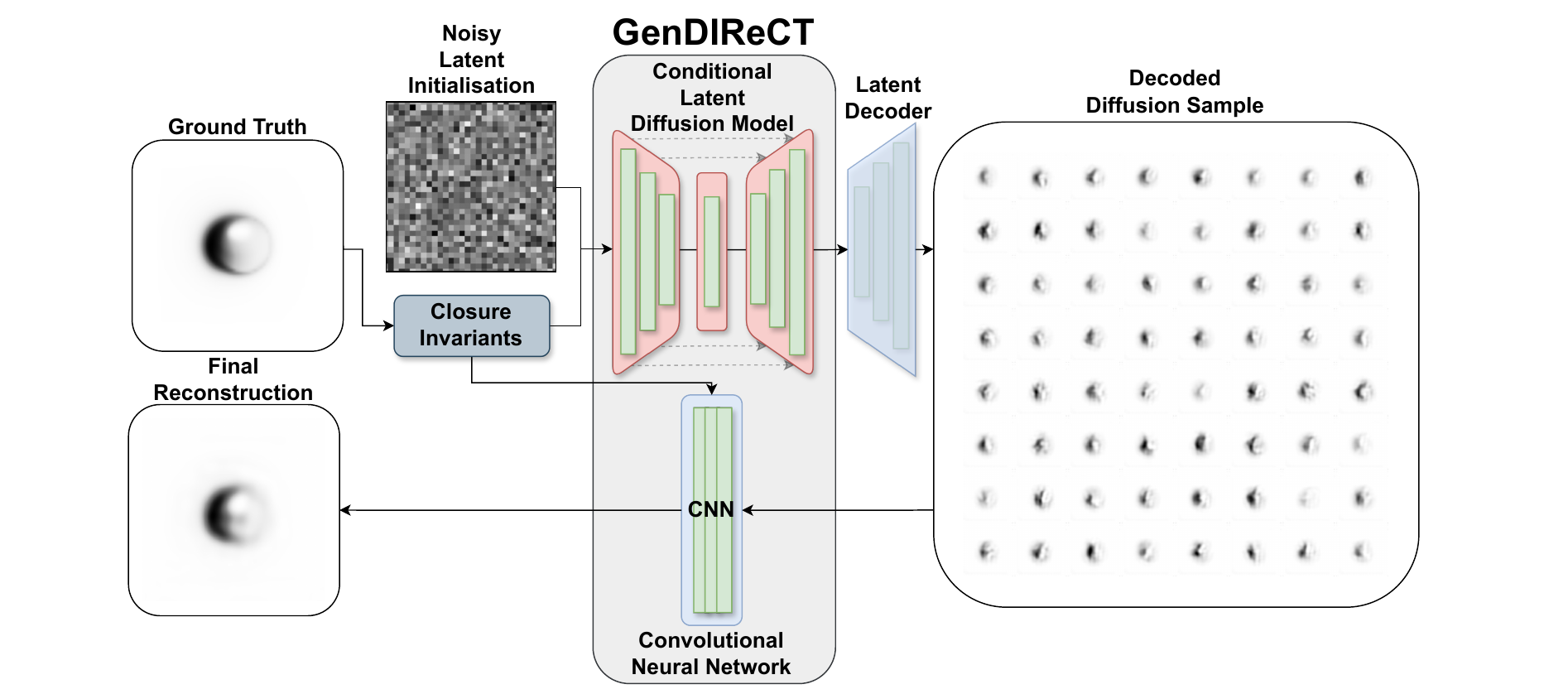}
    \caption[]{Overview of the \gendirect\ architecture, imaging sequence, and outputs. For a single reconstruction, the \gendirect\ model receives a dataset of closure invariants, which can also be derived from synthetic observation of a ground truth image. The pretrained conditional diffusion component samples images decoded from a distribution in the latent feature space, conditioned on the provided data. Subsequently, the convolutional neural network processes the diffusion sample and optimises for consistency with the input closure invariants, resulting in the final image reconstruction.}
    \label{fig:imaging-flowchart}
\end{figure*}

\section{\gendirect\ Imaging} \label{sec:imaging}
The Generative Deep learning Image Reconstruction with Closure Terms algorithm \citep[\gendirect;][hereafter referred to as \citetalias{Lai_2025_GenDIReCT}]{Lai_2025_GenDIReCT} is a novel imaging pipeline designed using the co-polar closure invariants \citep{Thyagarajan_2022_CI} to tackle the challenge of high-precision calibration and imaging from the sparse aperture coverage characteristic of VLBI data. Building on the earlier work of \direct, which showcased the potential of vision transformers in imaging \citep{Lai_2025_DIReCT}, \gendirect\ leverages the conditional image generation \citep[e.g.][]{Rombach_2021_StableDiff} of denoising diffusion probabilistic models \citep{DDPM_ho_2020} to effectively model the complex distribution of source morphologies conditioned on observed closure invariants. A separate multi-layer convolutional neural network (CNN) produces the final image reconstruction by optimising for the likelihood function containing the selected data fidelity metric. As the latent diffusion model is conditioned with closure invariants, it requires pretraining for each VLBI dataset, while the CNN is optimised for every individual reconstruction. We provide a simplified high-level illustration of the \gendirect\ architecture, imaging sequence, and its outputs in Figure \ref{fig:imaging-flowchart}, using a model black hole horizon radio image as an example. For a more detailed description of each individual component in the \gendirect\ architecture, their corresponding loss functions, and their training strategy, refer to \citetalias{Lai_2025_GenDIReCT}. 

Effectively, \gendirect\ replaces the prior and regularisation terms in regularised maximum likelihood-based methods with network-learned priors and regularisations, emulating image characteristics (such as positivity and smooth brightness distribution) from the non-astronomical image training dataset, primarily based on CIFAR-10 natural images \citep{Krizhevsky09_CIFAR10}. Moreover, the generative nature of \gendirect\ enables the imaging pipeline to potentially capture degeneracies and multimodalities in the solution space, by illustrating clusters of candidate solutions for each dataset. However, the integration of the CNN image refinement component does not fit within a Bayesian interpretation; therefore, clusters of \gendirect\ image reconstructions do not directly represent image reconstruction uncertainty.

In the \citetalias{Lai_2025_GenDIReCT} introductory paper, the performance of \gendirect\ was quantitatively evaluated on synthetic datasets derived from the EHT array, achieving excellent performance on data and image metrics for both trained and untrained source morphologies. When tested on the Next Generation Event Horizon Telescope (ngEHT) total intensity analysis challenge, \gendirect's performance was competitive with other state-of-the-art image reconstruction algorithms, while not requiring accurate calibration nor hyperparameter fine-tuning. Based on its performance on synthetic datasets, it's expected that \gendirect\ can be an effective blind imaging algorithm on public EHT data, offering an independent constraint on source morphology. However, until this work, \gendirect\ has not been tested on real data from the EHT.

For the first time, we adapt \gendirect\ to real datasets, by modelling the aperture coverage of the calibrated frequency-averaged 3C~279 and Centaurus~A data from the \hops\ pipeline \citep{Blackburn_2019_EHTHOPS}, where the visibility data is averaged across $\sim5$-minute scans prior to computing the closure invariants. Models are trained separately for each observing day on identical image datasets containing augmentations of both natural images and simple shapes, as described in \citetalias{Lai_2025_GenDIReCT}. Unless otherwise stated, the default field-of-view for 3C~279 is $225\times225\,\mu{\rm as}^2$ and for Centaurus~A a $450\times450\,\mu{\rm as}^2$ field-of-view was selected. All reconstructed images are produced on a $64\times64$ pixel grid, which is predetermined by the \gendirect\ architecture. For a discussion on alternative fields-of-view and their effect on the image reconstruction, refer to \ref{appendix:fov-effects}.

Compared to the description of the imaging sequence described in \citetalias{Lai_2025_GenDIReCT}, the differences implemented in this study are that we have enlarged the diffusion sample entering the convolutional neural network component by four times, enabling a wider and more diverse sampling of the latent parameter space. To facilitate the larger input, we extended the maximum number of training epochs to allow the convolutional neural network to converge. All other aspects, including the training, optimisation, loss functions, and network architecture remain unchanged from \citetalias{Lai_2025_GenDIReCT}. 

\subsection{Closure invariants}
\gendirect\ utilises co-polar interferometric closure invariants \citep{Thyagarajan_2022_CI} to condition the generative latent diffusion model and optimise the CNN image refinement model. Under the Abelian gauge theory formalism, closure invariants are constructed from elementary triangular plaquette variables called ``advariants'', which are defined for any pair of array elements $(a,b)$ pinned on a fixed reference vertex indexed at 0,
\begin{equation}
    \mathcal{A}'_{\rm{0ab}} = \mathcal{V}'_{0a}(\mathcal{V}'^{*}_{ab})^{-1}\mathcal{V}'_{b0} = \abs{g_0}^2\mathcal{A}_{\rm{0ab}}\,,
\end{equation}
where $\mathcal{V}_{ab}$ are the corrupted visibilities from the baseline between stations $a$ and $b$, $0$ is the reference station, and $\abs{g_0}^2$ is an unknown scaling factor identical on all complex advariants associated with the reference station. The complete and independent set of calibration-independent closure invariants can be constructed by normalising all advariants in each scan by their $L_2-$norm (other options are possible), resulting in $N_\textrm{s}^2 - 3N_\textrm{s} + 1$ real-valued closure invariants. Throughout this work, we have selected the highest sensitivity station in each scan (often ALMA when available) as the reference station, following the statistical analysis of \citet{Blackburn_2020} which proved that constructing triangular closure phases centered around baselines of the most sensitive station station would minimise the closure phase covariances when the array's sensitivity is dominated by a single station, such as in the case of the EHT with ALMA. However, a different selection for the reference vertex station would result in a new set of closure invariants carrying equivalent information. We explore the effect of an alternative selection of reference station on \gendirect\ image reconstructions in \ref{appendix:baseid}.

Closure invariants carry identical calibration-independent information as traditional closure phases and closure amplitudes. However, the Abelian gauge theoretic framework for deriving closure invariants is a unified formalism where measurements are derived entirely from independent triads, the simplest non-trivial loop. This carries some advantages compared to the traditional formalism, which necessitates separate treatment for the closed triangular and quadrilateral station loops. Further discussion of the co-polar closure invariants formalism and its advantages are described in \citet{Thyagarajan_2022_CI} with the polarimetric non-Abelian extension in \citet{Samuel_2022}.

\subsection{Model validation}
Prior to applying each of the trained \gendirect\ models on their corresponding EHT data, we quantify the performance of each model on a diverse set of synthetic datasets. Each of the validation tests can be categorised into one of two types: data corruption and image tests. In the data corruption category, we test the performance of the model under the effects of two additional sources of noise that are not removed by the closure invariants construction: thermal noise and baseline-dependent errors, as well as data corruption coming from an extended source with emission outside the reconstruction field-of-view. In the image test category, we evaluate the performance of the model on basic geometric shapes, selected images from \imagenet\ \citep{deng2009imagenet}, the standard \mnist\ handwritten digits \citep{deng2012mnist}, and astrophysically relevant images. On each of the challenges, we quantify the performance of the model on the following metrics, which evaluate both the data and image reconstruction fidelity.

\subsubsection{Fidelity metrics}
In this work, we employ the reduced $\chi^2_{\rm{CI}}$ metric to evaluate data consistency between different image reconstructions. The reduced $\chi^2_{\rm{CI}}$ is defined as the sum of error-normalised square residuals, divided by the number of independent data terms,
\begin{equation}
        \chi^2_{\rm{CI}} = \frac{1}{N_{\rm{ci}}}\sum_{i}^{N_{\rm ci}}\left[\frac{\left(\mathcal{C}(A)_i - \mathcal{C}({B})_i\right)^2}{\sigma_i^2}\right]\,, \label{eq:cnn-loss}
\end{equation}
where the operator $\mathcal{C}$ maps an image to its closure invariants, indexed by $i$, for a fixed observation arrangement, and $N_{\rm ci}$ is the total number of independent closure invariants. The normalisation, $\sigma_i$, is the expected uncertainty of the $i$-th closure invariant, which is sensitive to the noise model and observation details. Where the ground truth image is unknown, $\mathcal{C}(B)$ is replaced with the observed data, which is compared to $\mathcal{C}(A)$ for the reconstruction, $A$. Different decisions, such as the reference station selection or data aggregation strategy, can impact the size of the resulting set of closure invariants and their covariances. As such, in all cases throughout this paper, when $\chi^2_{\rm{CI}}$ is measured with EHT data, we use the public \hops-reduced data with no additional post-processing. This ensures that all of the $\chi^2_{\rm{CI}}$ data metrics reported for each dataset are evaluated on a consistent metric, regardless of the data processing decisions. 

While thermal errors are relatively straightforward to propagate from each radio station's system equivalent flux density (SEFD), other sources of systematic errors, such as residual calibration offsets, time-dependent errors from long coherent averaging of visibilities, and limitations of a linear error propagation model, can also affect both visibilities and closure invariants. In some cases and particularly at high SNR, the systematic uncertainty can become dominant over more easily quantifiable thermal noise. In this work, we adopt a relatively conservative homogeneous systematic uncertainty estimate of $5\%$ in addition to the measured thermal noise, which is on the same order as the measured inter-pipeline differences and aligned with the recommended systematic error budget of the 3C~279 and M87 data \citep{EHT_2019_Data}.

In validation tests where the ground truth is known and for relative comparisons between image reconstructions, we introduce the maximum normalised cross-correlation metric, $\rho_{\rm{NX}}$, to evaluate the image fidelity or correspondence between reconstructions independently from the data fidelity metric. The maximum normalised cross-correlation metric is defined as,
\begin{equation}
    \rho_{\rm{NX}}(A, B) = \frac{1}{M}\max\abs{{\mathcal{F}^{-1}\{\mathcal{F}\{\hat{A}\}\mathcal{F}\{\hat{B}\}^*\}}}\,,
\end{equation}
where $A$ and $B$ are images and the hat operator, $\hat{I} = (I - \bar{I})/\sigma_I$, is an image, $I$, normalised by its self-evaluated mean and standard deviation. Operations $\mathcal{F}$ and $\mathcal{F}^{-1}$ represent the forward and inverse Fourier transforms, respectively. By taking the maximum value, $\rho_{\rm{NX}}$ is independent of image translations and changes in the total flux level, which is ideal for evaluating image reconstructions from closure information where these properties are unconstrained. We also employ the $\rho_{\rm{NX}}$ metric to determine the corresponding image translation for aligning separate image reconstructions by their maximum cross-correlation. We add that when estimating the relative correspondence between reconstructions, both images would be blurred by the fitted uniformly-weighted clean beam prior to computing $\rho_{\rm{NX}}$. We choose to compare $\rho_{\rm{NX}}$ after a final convolution with the nominal image resolution due to the heterogeneous origin of EHTC reference reconstructions, where the Centaurus A reference is a single refined image reconstruction, while the 3C~279 reference is a combined reconstruction from three separate imaging algorithms. Generally, in this work, we represent the \gendirect\ output with the median reconstruction, which is an aggregated representation of numerous individual reconstructions, similar to the latter case of 3C~279. By smoothing out high-resolution features, such as those present in the EHTC reference image of Centaurus A, with the nominal image resolution, we can construct a more equitable comparison between image reconstruction methods using the $\rho_{\rm{NX}}$ correspondence metric.

\subsection{Performance on validation tests} \label{sec:validation-performance}
In this section, we summarise the performance of two models on both image and data corruption tests: one model is trained with the 11 April 2017 dataset of 3C~279 to produce images on a $225\times225\,\mu{\rm as}^2$ field-of-view and the other model is trained with the Centaurus~A dataset on a $450\times450\,\mu{\rm as}^2$ field-of-view. Note that the 11 April 2017 dataset of 3C~279 has the most complete aperture coverage of the four observation dates. Across all validation tests, we find that both models can consistently achieve $\chi^2_{\rm{CI}} < 2$ on the data fidelity metric, except in cases of severely corrupted data and significant flux outside the reconstruction field-of-view. 

First, we describe the performance of both \gendirect\ models on data corruption tests, which include scaling the thermal noise, the addition of baseline-dependent errors, and an extended source with emission outside the reconstruction field-of-view. Data corruption validation tests are performed with synthetic data on a simulated event-horizon-scale image with a total compact flux density of $1$ Jy. The choice of image can shift the nominal performance of the \gendirect\ models; thus, we place no emphasis on the relative performance between the trained models and examine only the trends of a single model with different levels of noise. On thermal noise tests, we scale the thermal noise from $0.01 - 100\times$ the standard SEFD, and evaluate both image and data fidelity by the $\rho_{\rm{NX}}$ and $\chi^2_{\rm{CI}}$. In all cases, the expected $\sigma^2$ normalisation of each $\chi^2_{\rm{CI}}$ measurement is based on the standard SEFD level for a consistent comparison. We find that all models retain consistent performance on $\rho_{\rm{NX}}$ up to $\sim3\times$ SEFD; after which, $\rho_{\rm{NX}}$ indicates that the reconstruction is unreliable. Correspondingly, the models' performance on $\chi^2_{\rm{CI}}$ worsen gradually as noise level increases, passing $\chi^2_{\rm{CI}}\sim1$ between $\sim3-30\times$ standard SEFDs. 

We define the baseline-dependent errors as an additional zero-mean Gaussian corruption applied to each baseline, with its standard deviation defined as some fraction of the visibility amplitude. Like the thermal noise test, we find consistent performance in $\rho_{\rm{NX}}$ and gradually degraded performance on $\chi^2_{\rm{CI}}$. The critical threshold for image fidelity is when the baseline noise reaches $10\%$ of the visibility amplitude, while the data fidelity passes $\chi^2_{\rm{CI}}\sim1$ at $30\%$. For the validation test on the extended source with emission beyond the reconstruction window, we first construct a crescent within the field of view and define a large-scale disk with a diameter $2\times$ the size of the reconstruction window width and scale the relative total flux between the two components. In all models, we find that \gendirect\ maintains consistent performance on both image and data metrics as the relative flux of the large-scale disk component grows until the flux ratio becomes unity. Once the extended disk becomes the dominant component, it's not possible for \gendirect\ to reconstruct the disk within its limited field-of-view, which significantly degrades its performance on all metrics. Therefore, as long as the majority of flux is contained within the limited reconstruction window, \gendirect\ can produce an accurate image reconstruction. 

Image validation tests target both general out-of-distribution morphologies, such as \mnist\ digits and \imagenet, as well as basic geometric shapes and astrophysically relevant images. All synthetic data for these tests are created with thermal noise from standard SEFD values. As we have previously established that \gendirect\ has minimal difficulty fitting the $\chi^2_{\rm{CI}}$ data metric for standard SEFDs, we focus primarily on the image fidelity metric for this discussion. We first note that the test on basic geometric shapes, which includes Gaussians, rings, crescents, and doubles, establishes a baseline performance for each model as it tests the trained model's performance on structures similar to those in its training dataset. On this test, we note that all models perform well with image fidelity scores reaching $\rho_{\rm{NX}} \sim 0.95$. We then test the models on the out-of-distribution sources, such as handwritten digits from the \mnist\ dataset, natural images from \imagenet, and a variety of astrophysically relevant images, which include images of a protoplanetary disk, protostellar cluster, protostellar envelope, M51 H$\alpha$, g41 continuum, 3C288 radio galaxy, and 30Dor in the near-infrared. All images are collected from the CASA data-bank\footnote{\href{https://casaguides.nrao.edu/index.php/Sim_Inputs}{https://casaguides.nrao.edu/index.php/Sim\_Inputs}} and resized to the field-of-view of the trained model. While some individual reconstructions reach $\rho_{\rm{NX}} \sim 0.95$, most reconstructions fall within the range of $\rho_{\rm{NX}} \sim 0.8 - 0.9$. Out of the three datasets, \gendirect\ performs the best on \mnist\ due to its clean zero-value background. In summary, across all tests, the trained \gendirect\ models are performing within expectations, indicating that the closure invariant datasets are sufficiently constraining for a variety of possible image morphologies.

\begin{figure*}
    \centering
    \begin{subfigure}{0.95\textwidth}
    \centering
    \includegraphics[width=0.95\textwidth]{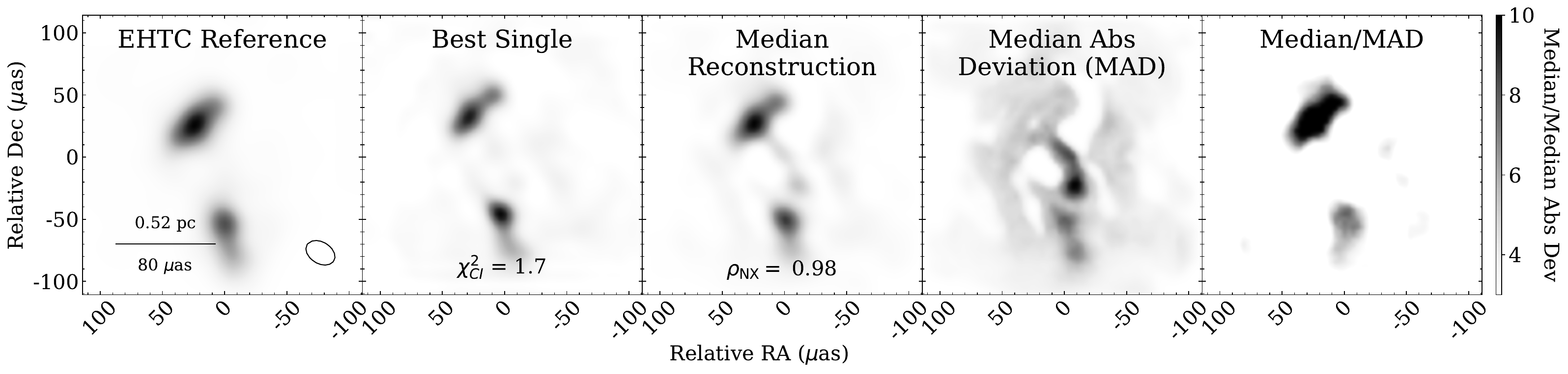}
    \end{subfigure}
    \begin{subfigure}{0.95\textwidth}
    \centering
    \includegraphics[width=0.95\textwidth]{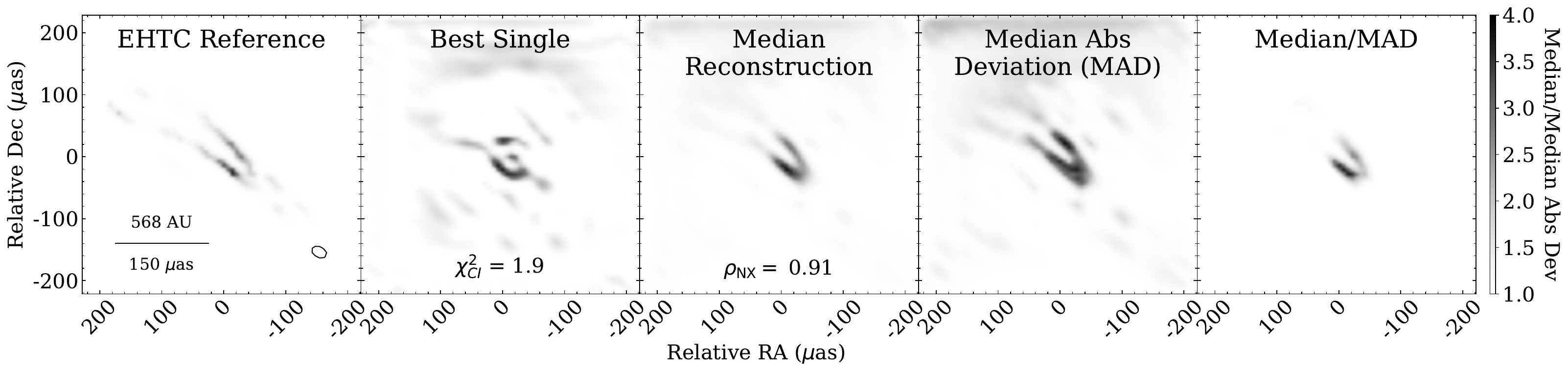}
    \end{subfigure}
    \caption[]{Comparison between the EHTC reference reconstructions of blazar 3C~279 (top) and Centaurus~A (bottom) with their corresponding \gendirect\ reconstructions. The dataset used in the reconstruction of 3C~279 is from 11 April 2017. \gendirect\ outputs include the best single reconstruction as measured by the reduced $\chi^2_{\rm{CI}}$, the median reconstruction, the median absolute deviation, and ratio images. The ratio image displays the median image over the median absolute deviation, illustrating reconstruction confidence. The synthesised beam is visualised in the left panel with an elliptical model. All 3C~279 reconstructions are plotted on a $225\times225\,\mu{\rm as}^2$ field-of-view, while Centaurus~A reconstructions are pictured on a $450\times450\,\mu{\rm as}^2$ field-of-view. All \gendirect\ reconstructions (and the EHTC Centaurus A reference image) are presented without clean beam convolution, while the EHTC reference for 3C 279 has previously been convolved with a $20\,\mu$as circular beam.}
    \label{fig:3C279CenA_compare}
\end{figure*}

\section{Results} \label{sec:results}

With the \gendirect\ models trained and validated as described in Section \ref{sec:validation-performance}, we present our results on 3C~279 (top) and Centaurus~A (bottom) in Figure \ref{fig:3C279CenA_compare}. For each dataset, we run \gendirect\ for 66 GPU-hours on the NVIDIA-H100, which produces several hundred image reconstructions ($\sim300$), although the exact number varies by $\sim10\%$ depending on the dataset. The leftmost panels in each row contain the reference reconstructions from the EHTC \citep{Kim_2020_3C279, Janssen_2021_CenA}, followed from left-to-right by the best single reconstruction from the \gendirect\ pipeline as measured by the minimum value of closure invariants reduced $\chi^2_{\rm{CI}}$, the median of all \gendirect\ reconstructions on the respective dataset aligned by $\rho_{\rm{NX}}$, the median absolute deviation (MAD), and the ratio image between the median image and the median absolute deviation. The maximum normalised cross-correlation between the median \gendirect\ reconstruction and the EHTC reference reconstruction is shown on the median panel. We note that although we use the aligned median reconstruction throughout this study to aggregate the generative results of \gendirect\ into a single image, the median is not intended to reflect the best-performing image in metrics $\chi^2_{\rm{CI}}$ or $\rho_{\rm{NX}}$. However, it's likely the most suitable image for comparison with the EHTC references, which are likewise informed by multiple pipelines under different imaging conditions, aligned, and averaged \citep{Kim_2020_3C279, Janssen_2021_CenA}. The right-most ratio panel in each row illustrates a `signal-to-noise' ratio of the reconstruction with darker regions reflecting more consistently reconstructed features.

The top row of Figure \ref{fig:3C279CenA_compare} corresponds to reconstructions of 3C~279 in a $225\,\mu{\rm{as}}\times225\,\mu{\rm{as}}$ field-of-view, specifically reconstructions resulting from the 11 April 2017 dataset, which is the dataset with the best single-day aperture coverage \citepalias{Kim_2020_3C279}. The median of \gendirect\ reconstructions exhibits a $\rho_{\rm{NX}} = 0.98$ correspondence with the EHTC reference reconstruction, though there are apparent differences in the significance of the flux bridging the assumed core (northern component) and ejecta (southern component), as well as in the presence of more spurious emission on both the east and west sides. While some flux is marginally detectable in the EHTC reference reconstruction between the bright components, it is significantly fainter than in the \gendirect\ median reconstruction. However, as shown in the median/MAD ratio image, the reconstruction confidence is low in both the emission bridge and the flanking emission compared to the two bright components. 

In the bottom row of Figure \ref{fig:3C279CenA_compare}, the median \gendirect\ reconstruction of Centaurus~A has a maximum normalised cross-correlation score of $\rho_{\rm{NX}} = 0.91$ when compared with the EHTC reference model reconstruction. While the highest surface brightness components of the edge-brightened approaching jet were reproduced by the \gendirect\ reconstruction, the counter-jet emission and some of the fainter features extending out to angular distances of $\sim 200\,\mu$as from the jet apex are faint or absent. Instead, spurious emission that is not well-constrained by the data can be observed near the top of the reconstruction window in some of the \gendirect\ reconstructions. The normalised cross-correlation score of Centaurus~A relative to the EHTC reference reconstruction was estimated after clipping the spurious emission near the top of the \gendirect\ image frame in order to compare the morphologies of the central compact flux distribution. According to the ratio image, the only consistently reconstructed features from \gendirect\ are the two bright ridge-lines along the sheath and near the base of the approaching jet and the overall reconstruction confidence is lower than that of 3C~279. 

In \ref{appendix:alternative-imaging}, we explore the effects of different data processing and pre-imaging considerations, such as variations in the field-of-view, alternative data aggregation strategies, and a different selection of the closure triad reference station, on the resulting \gendirect\ image reconstruction. We show that the image reconstructions shown in this section are robust against sensible modifications to the reconstruction window and data processing methods. Throughout this section, image reconstructions from both selections of the reference station contribute to the \gendirect\ reconstruction sample, but reconstructions from alternative fields-of-view or with averaging closure invariants are excluded. 

\subsection{Reconstruction multimodalities}
\begin{figure*}
	\includegraphics[width=0.92\textwidth]{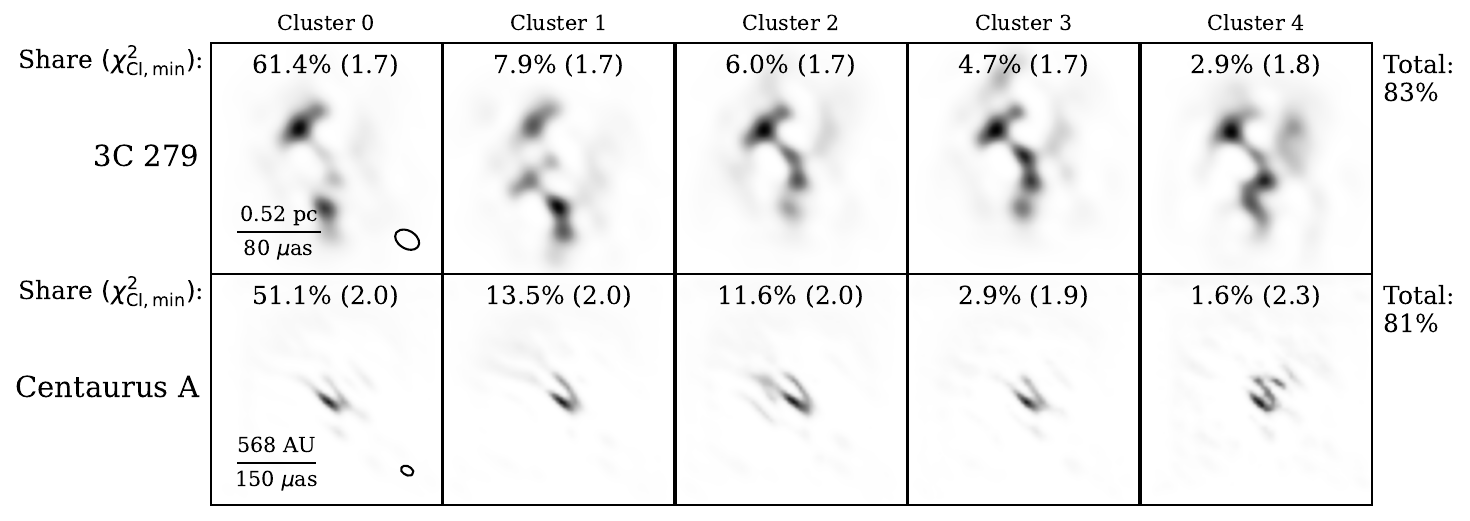}
    \caption[]{Reconstruction clusters for the EHT observations of 3C~279 (top row) and Centaurus~A (bottom row). Each cluster is represented by the median image and we show the percentage share alongside the minimum $\chi^2_{\rm{CI}}$ (in parenthesis) of all reconstructions identified with that cluster. The first image of each row shows the linear scale and fitted clean beam.}
    \label{fig:clusters}
\end{figure*}

As a consequence of our generative approach, we are able to identify data-consistent, but perceptually diverse multimodalities in the resulting image reconstruction, similar in effect to normalising flow techniques \citep[e.g.][]{Feng_2024} or evolutionary algorithms \citep[e.g.][]{Muller_2023_MOEAD}. Clusters of similar reconstructions are identified by the Hamming distance in the discrete cosine transform hash, as described in \ref{appendix:phashing}. Figure \ref{fig:clusters} illustrates five of the most populated image reconstruction clusters of 3C~279 (top row) and Centaurus~A (bottom row), where each cluster is represented by its median image and displayed in descending order (left-to-right) by the share of all reconstructions belonging to the cluster. In identifying clusters of Centaurus~A reconstructions, we have applied a Gaussian filtered circular mask to remove some spurious low surface brightness emission typically located near the image periphery, as observed in Figure \ref{fig:3C279CenA_compare}.

While the image clusters of 3C~279 all show at least two extended components with a consistent relative angular separation, the clusters differ in the brightness ratio between the assumed core and ejecta, as well as in the number and flux of any additional components in the intervening space. The dominant cluster with $>60\%$ share of all reconstructions is the most morphologically consistent with the EHTC reference reconstruction, while several of the alternative solution clusters, such as Clusters 2, 3, and 4, exhibit brighter emission trailing towards the SW direction relative to the core. Often these additional components are accompanied by enhanced faint emission in the western direction of the reconstruction window, from which we can infer that any sparsity regularisation would heavily penalise such solutions. 

The dominant cluster for Centaurus~A is comprised of a $>50\%$ share of all reconstructions. Alternative solution clusters primarily differ from the dominant cluster in the brightness ratio between the two approaching jet ridges and the low-luminosity extension of the jet stream, although most solutions have the southern jet sheath as the brighter feature. Certain clusters, such as Cluster 1 for 3C~279 as well as in Clusters 2 and 4 for Centaurus~A, exhibit phantom repetition of similar structures. These secondary phantom images are not unusual in the VLBI context, as they are symptoms of the poor aperture coverage and phase error \citep[i.e.][]{Chael_2018_ehtim, Muller_2023_MOEAD}. 

We emphasise that while the \gendirect\ approach can be used to identify and explore clusters of of candidate reconstructions, the process is not strictly identical to sampling from a posterior distribution in a Bayesian framework. Rather, the diffusion model of \gendirect\ is trained to emulate the expected conditional probability distribution of its training data in the latent domain. Then the convolutional neural network further optimises an aggregated image reconstruction from the diffusion-generated sample according to the error-weighted measurements. While the generative output of this process contains clusters of candidate image reconstructions, the image distribution does not allow for quantification of statistical uncertainty.

\section{Discussion} \label{sec:discussion}

As an imaging algorithm conditioned on only closure information, the results of \gendirect\ are insensitive to different sets of calibration assumptions or residuals in the calibration solution, enabling an independent verification of image morphologies obtained through VLBI methods. Moreover, absent of explicit priors, initialisations, and regularisations, \gendirect\ provides a relatively simple and reproducible imaging tool for VLBI datasets, which is also robust against several different pre-imaging considerations as shown in \ref{appendix:alternative-imaging}. With only minimal and sensible modifications to the size of the diffusion sample and maximum number of epochs for convergence, we have applied the methodology and imaging sequence of \citetalias{Lai_2025_GenDIReCT} on real EHT data for the first time in this work. 

In the previous section, we presented the image reconstructions of the \gendirect\ model trained on the 11 April 2017 dataset of 3C~279 and 10 April 2017 dataset of Centaurus~A. We illustrated the multimodalities in the resulting image reconstructions by clustering candidate solutions and presented an aggregated median reconstruction, alongside the median absolute deviation, and ratio images. We measured the best single reconstruction data fidelity $\chi^2_{\rm{CI}}$ and maximum cross-correlation, $\rho_{\rm{NX}}$ with the EHT reference images, finding consistent results with the published images. In this Section, we present a more detailed morphological comparison between the \gendirect\ reconstructions and the EHTC images. Additionally, we study the other EHT datasets of 3C~279 taken on different days in April 2017, and evaluate the inter-day variability through independent \gendirect\ image reconstructions. 

\subsection{Morphological comparison} \label{sec:EHT-morpho-comparison}

\begin{figure}
	\includegraphics[width=0.95\columnwidth]{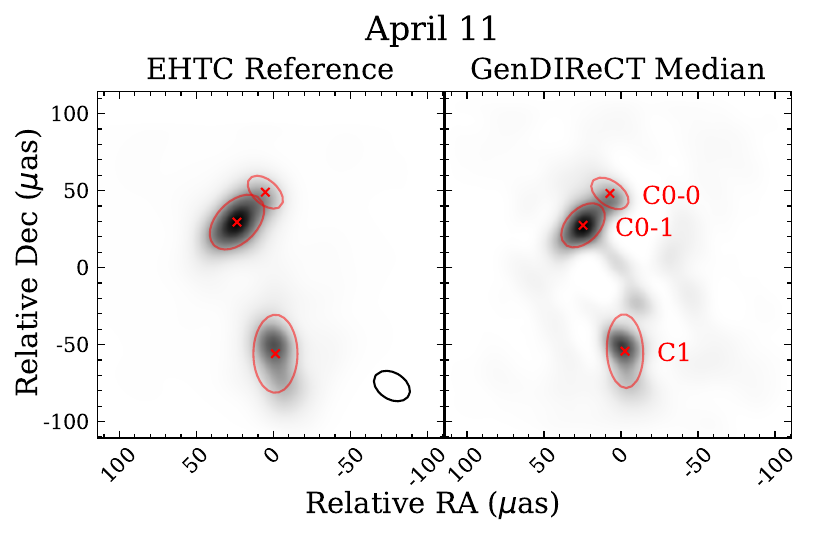}
    \caption[]{Illustration of the three elliptical Gaussian component fits of both the EHTC reference (left) and median \gendirect\ (right) reconstruction on the 11 April 2017 dataset of 3C~279. The centroid of each component is marked with an 'x' and a single contour is drawn at the half-amplitude. }
    \label{fig:3C279-fitted}
\end{figure}

We have shown in Figure \ref{fig:3C279CenA_compare} that the \gendirect\ median reconstructions of both 3C~279 and Centaurus~A are rated at $\rho_{\rm{NX}} > 0.9$ on the maximum normalised cross-correlation metric when compared to the EHTC reference reconstructions. In this section, we quantitatively compare measurable morphological characteristics between the EHTC and \gendirect\ reconstructions by applying identical parameterised fitting algorithms to both images. 

We jointly fit the reconstructions of 3C~279 with three elliptical
Gaussian components: two for the extended core and a single component for the ejecta, applying the Levenberg-Marquardt algorithm for parameter estimation.
We adopt the nomenclature used in \citetalias{Kim_2020_3C279} to label each of the components. Note that while \citetalias{Kim_2020_3C279} fits 6 Gaussian components in the visibility model-fitting analysis, we combine the C1-0, C1-1, and C1-2 components into a single C1 component, and do not attempt to fit C0-2, which is faint and likely undetectable in the \gendirect\ reconstruction. Figure \ref{fig:3C279-fitted} illustrates the three fitted components on both the \citetalias{Kim_2020_3C279} reference and median \gendirect\ reconstructions, with the centroid of each component marked by an `x' and contour drawn to illustrate the Gaussian half-amplitude. We measure the intensity ratio between C0-1 and C1 ($I({\textrm{C0-1}})/I({\textrm{C1}})$), separation ($\abs{\textbf{C0} - \textbf{C1}}$), and angle ($\theta_{\textbf{C1} - \textbf{C0}}$) between the flux weighted C0 and C1, as well as the angular extension of the C0-1 component (PA$_{\textbf{C0-1}}$). The measured quantities are tabulated and compared in Table \ref{tab:fitted-params}. Note that $\theta_{\textbf{C1} - \textbf{C0}}$ and PA$_{\textbf{C0-1}}$ are reported in degrees north of west. We find that all measured quantities to be statistically similar within $\sim1\sigma$ uncertainty intervals, except for the angular separation between C1 and the flux-weighted centroid of C0. However, the angular discrepancy is $\sim0.7\,\mu\rm{as}$, which is insignificant compared to the clean beam semi-major axis ($\sim25\,\mu{\rm{as}}$) and the pixel scale of the \gendirect\ reconstruction ($\sim3.5\,\mu{\rm{as}}/\rm{pixel})$. If one considers adding in quadrature a systematic position error of 5\% the clean beam geometric mean half-width at half-maximum, the angular separation discrepancy is no longer statistically significant. 
Altogether, these results effectively demonstrate that the robustly detected core and ejecta components of 3C~279 are morphologically similar within expected uncertainties between the EHTC reference and median \gendirect\ reconstruction, which is supported by the image correspondence metric, $\rho_{\rm{NX}} = 0.98$. 

\setlength{\extrarowheight}{3pt}
\begingroup
\begin{table}
\caption {\label{tab:fitted-params} Tabulated quantities comparing fitted morphological parameters for both 3C~279 and Centaurus~A.} 
\begin{tabular}{llcc}
\hline \hline
 & Parameter & EHTC Reference & \gendirect\ Median  \\
 \hline
3C~279 & $I({\textrm{C0-1}})/I(\textrm{C1})$ & $1.52 \pm 0.01$ & $1.53 \pm 0.02$ \\
\citepalias{Kim_2020_3C279} & $\abs{\textbf{C0} - \textbf{C1}} \, [\mu \textrm{as}]$ & $90.9 \pm 0.2$ & $90.2 \pm 0.2$ \\
& $\theta_{\textbf{C1} - \textbf{C0}}$ [deg$^{*}$] & $-75.7 \pm 0.1$ & $-75.6 \pm 0.1$ \\ 
& PA$_{\textbf{C0-1}}$ [deg$^{*}$]& $44.5 \pm 0.5$ & $44.2 \pm 0.7$ \\ \hline
CenA & $R_{\rm{s/n}}$ & $1.5\pm0.2$ & $1.4\pm0.1$ \\
\citepalias{Janssen_2021_CenA} & $\theta_{\rm{s/n}}$ [deg] & $12.2 \pm 0.1$ & $12.3 \pm 0.3$ \\ 
\hline \hline
\multicolumn{4}{l}{$^{*}$\footnotesize Measured north of west.}
\end{tabular}
\end{table}
\endgroup
\setlength{\extrarowheight}{0pt}

\begin{figure}
\includegraphics[width=0.95\columnwidth]{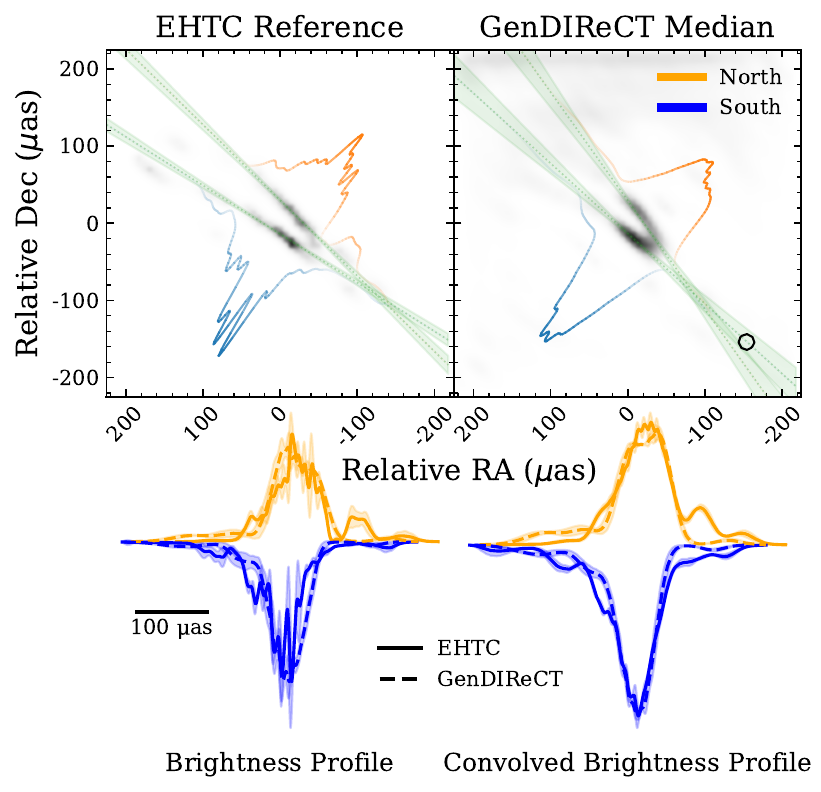}
    \caption[]{Illustration of the EHTC reference (top-left) and median \gendirect\ (top-right) reconstructions fitted with straight lines for each of the northern and southern edge-brightened jet ridgelines. The surface brightness along the fitted ridgeline is represented by the distribution plotted orthogonally to each line. In the bottom row, we compare the northern and southern surface brightness profiles between the EHTC (solid line) and \gendirect\ (dotted line) reconstructions. The bottom-right panel compares the profiles for both reconstructions after convolving with a common $20\,\mu$as circular beam, which is illustrated on the \gendirect\ median image.}
    \label{fig:CenA-fitted}
\end{figure}

For Centaurus~A, we jointly fit each of the edge-brightened ridges with 
separate linear profiles and consider the intensity ratio between the north-south ridgelines as well as the opening angle, which we define as the angle formed between the two fitted lines. We show both the \citetalias{Janssen_2021_CenA} reference and \gendirect\ image reconstructions fitted with the double linear model in Figure \ref{fig:CenA-fitted}, alongside a comparison of the brightness profiles normalised by the peak flux. The linear fits are performed on both reconstructions blurred by a common 20 $\mu$as circular beam. From the fitted model, we measure the brightness asymmetry ($R_{\rm{s/n}}$) between the two jet ridges and the opening angle ($\theta_{\rm{s/n}}$), tabulating the results in Table \ref{tab:fitted-params}. We find no statistically significant discrepancies between the two fitted models on either quantity; however, we note that while the opening angle is near-identical, the two fitted line models appear to be rotated by $\sim10 \pm 4$ deg with respect to each other. This is likely caused by the extended lower-luminosity features located further away ($\sim200\,\mu$as) from the jet apex, which are present in the EHTC reference image, but significantly fainter or altogether absent in the \gendirect\ reconstruction. The extended emission drives the fitted model slope shallower in the EHTC reference reconstruction compared to the median \gendirect\ reconstruction. We also find the Centaurus~A jet to exhibit more high-frequency structure in the \citetalias{Janssen_2021_CenA} jet model, but differences in the brightness profiles in the proximity of the brightest components become insignificant once both images are convolved with a common 20 $\mu$as beam.

\subsection{3C~279 inter-day variability}

During the 2017 EHT observing campaign, 3C~279 was observed on four separate days between April 5 and 11. We chose to image 3C~279 with the April 11 data in Section \ref{sec:results}, as it is the day with the best single-day aperture coverage. However, \citetalias{Kim_2020_3C279} successfully imaged 3C~279 using all of the datasets in the 2017 observing campaign and robustly identified six compact bright features through a Gaussian model-fitting analysis \citep[\textsc{THEMIS};][]{Broderick_2020_Themis}. With the Gaussian component parameters, \citetalias{Kim_2020_3C279} observed prominent and rapid inter-day brightness variations, as well as relative proper motion between components of $\simeq 1\,\mu$as day$^{-1}$, corresponding to apparent superluminal velocities of $\beta_{\rm{app}} = \frac{v_{\rm{app}}}{c}\sim (13-15)\pm2$. 

\begin{figure*}
	\includegraphics[width=0.95\textwidth]{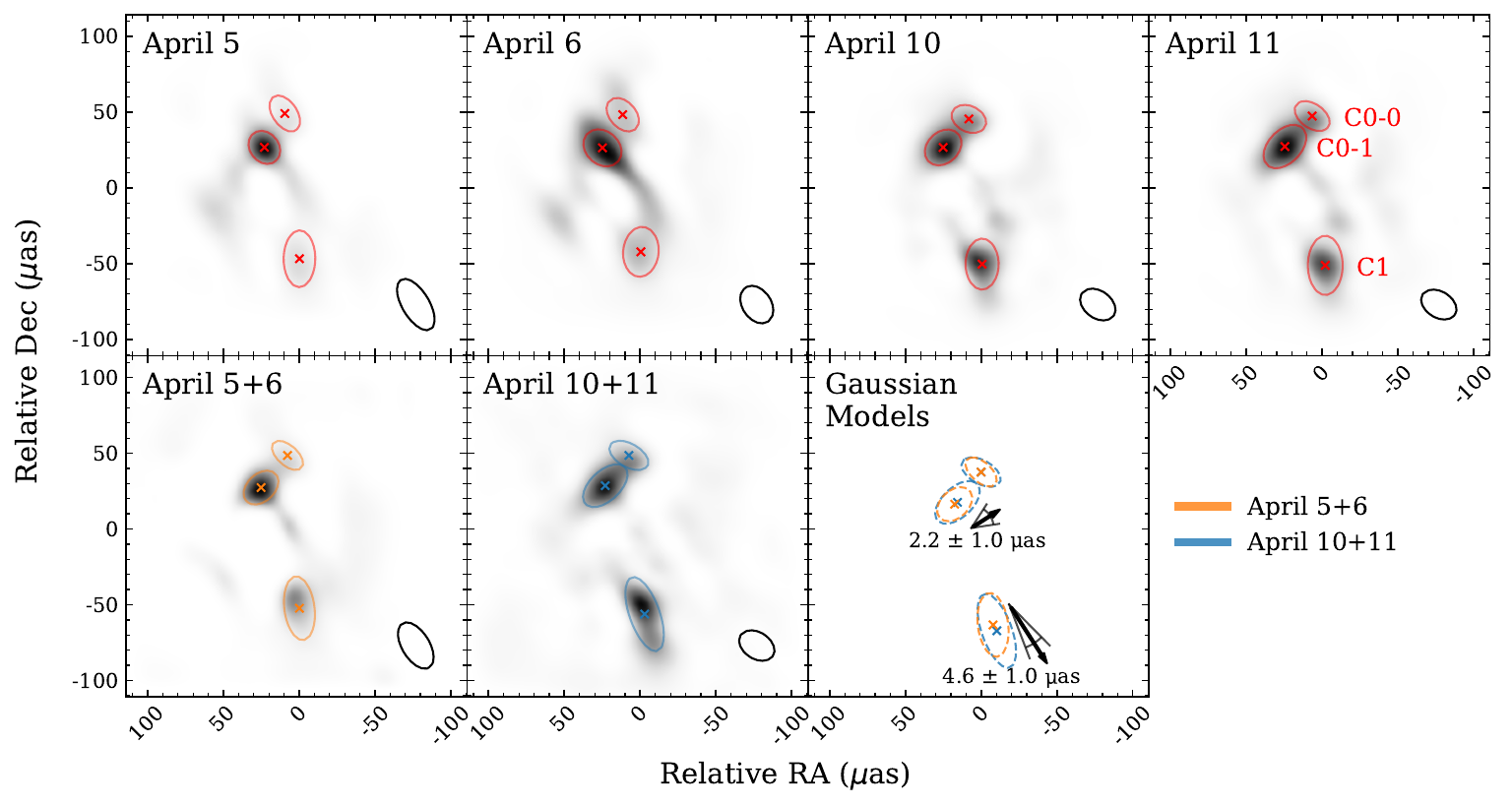}
    \caption[]{(Top row) Median \gendirect\ reconstructions of 3C~279 on each observation date. (Bottom row) Median \gendirect\ reconstructions using the combined dataset of the first and latter two observations dates. Each image is fitted with three Gaussian components, where the centroid is marked and a single contour is plotted at the half-maximum flux value. The components are labeled similarly to the designations from \citet{Kim_2020_3C279}. The final bottom-right panel plots the centroid and contours of the fitted Gaussian components in the combined dataset with C0-0 as the kinematic reference. Arrows point in the direction of proper motion and wedges illustrate the $1\sigma$ uncertainty in direction.}
    \label{fig:3C279-multiday}
\end{figure*}

In this section, we independently image the April 5, 6, and 10 datasets with \gendirect, presenting the median reconstructions in Figure \ref{fig:3C279-multiday} alongside the previously presented results on the April 11 dataset. Due to the more limited dynamic range of the closure-based reconstructions \citep[e.g.][]{Chael_2018_ehtim}, we identify three distinct emission components (C0-0, C0-1, and C1) as in Section \ref{sec:EHT-morpho-comparison}, each labeled accordingly with reference to \citetalias{Kim_2020_3C279}. Unlike the model-fitting analysis performed in \citetalias{Kim_2020_3C279} which fit Gaussian models to scan-averaged visibilities, we attempt a similar independent analysis in the image domain. We fit an elliptical
Gaussian to each of the three identified components in the median image, marking its centroid and contour at the half-maximum value in Figure \ref{fig:3C279-multiday}. On April 5, the C0-0 component is only be marginally detected. In this case, the Gaussian parameters are fit to the image with loose constraints from the detections on the other days. 

As shown in Fig. 1 of \citetalias{Kim_2020_3C279}, the aperture coverage on April 5 and 6 are notably more sparse than on April 10 and 11, particularly due to the absence of baselines involving the SPT, limiting the sensitivity in the North-South direction. Both images exhibit spurious emission features that are not clearly present in reconstructions from April 10 and 11 datasets. Meanwhile, both images from the later two days show a high degree of consistency. Following \citetalias{Kim_2020_3C279} in adopting C0-0 as the VLBI core and kinematic reference, we compare the motion of the centroids from the fitted C0-1 and C1-0 Gaussian components. When considering each day individually, the uncertainties in the fit parameters are too high to resolve proper motion on the order of 1 $\mu$as day$^{-1}$ in a $64\times64$ pixel grid covering $225\times225\,\mu{\rm as}^2$. Therefore, to increase the signal's significance, we train a \gendirect\ model on the concatenated datasets of April 5 and 6, as well as April 10 and 11, producing a median image reconstruction from each new combined dataset, and comparing the averaged differential motion across the approximately five-day separation. 

The bottom row of Fig. \ref{fig:3C279-multiday} presents the median \gendirect\ reconstructions on the combined datasets with the three-component fits. In the third panel on the bottom row, the fitted Gaussian components on the two separate epochs are overplotted with C0-0 set as the kinematic reference. We include a fiducial position uncertainty of 5\% the clean beam's geometric mean half-width at half-maximum as a systematic error added in quadrature to the measurement parameter uncertainties. For the April 5+6 and April 10+11 datasets, the additional systematic uncertainty on the measured centroids amounts to $\sim0.62\,\mu{\rm{as}}$ and $\sim0.53\,\mu{\rm{as}}$, respectively. Relative to C0-0, we detect a $2.2\pm1.0\,\mu$as proper motion in the C0-1 component toward the NW direction ($33\pm24$ deg) between the combined datasets. The proper motion measured for the C1 component is $4.6\pm1.0\,\mu$as in the SW direction ($-57\pm12$ deg), which is aligned with the large-scale jet direction \citep[e.g.][]{Fuentes_2023_RadioAstron}. Generally, the magnitude and direction of the proper motion estimates for C1 are consistent with those reported in \citetalias{Kim_2020_3C279} from the visibility model-fitting analysis, but the motion for our measured C0-1 sub-component is more consistent with that of \citetalias{Kim_2020_3C279}'s C0-2, though we don't attempt to model C0-2. The nearly perpendicular motions of \citetalias{Kim_2020_3C279}'s C0-1 and C0-2 complicate a direct comparison with our single C0-1 model. On the other hand, all of the C1 sub-components (C1-0, C1-1, and C1-2) are moving in a consistent direction in the \citetalias{Kim_2020_3C279} analysis. Therefore, we calculate that the motion of the flux-weighted C1 centroid in \citetalias{Kim_2020_3C279}'s analysis is $\Delta{\rm{RA}} = -1.2\pm 0.4\,\mu$as, $\Delta{\rm{Dec}} = -4.2\pm0.9\,\mu$as, with a total angular displacement of $4.4\pm0.9\,\mu$as. Altogether, these values are mutually consistent within $1\sigma$ with \gendirect's independent analysis, which measured $\Delta{\rm{RA}} = -2.5\pm1.0\,\mu$as, $\Delta{\rm{Dec}} = -3.9 \pm 1.0\,\mu$as, and total displacement of $4.6\pm1.0\,\mu$as, despite the fact that our analysis was performed on the image plane. We measure that the difference in the mean observation epoch between the combined two-day datasets is 5.39 days, with which we estimate the apparent proper motion of the C1 component to be $\beta_{\rm{app}} = 10\pm2$.

\section{Conclusion} \label{sec:conclusion}
\gendirect\ is a novel deep learning imaging algorithm designed to solve the ill-posed inverse problem of directly reconstructing images from a set of interferometric closure invariant measurables in the highly challenging VLBI context \citep{GenDIReCT_Zenodo, Lai_2025_GenDIReCT}. In previous work, the co-polar closure invariants formalism presented in \citet{Thyagarajan_2022_CI} and applied under the conditions of sparse aperture coverage was shown to be effective carriers of calibration-independent information on true source structure.

In this work, we present, for the first time, the application of the \gendirect\ imaging technique on real data from the Event Horizon Telescope on active galactic nuclei.  The two targets in this work are the archetypal blazar 3C~279 and the radio galaxy Centaurus~A, observed by the EHT in 2017. Below, we summarise the main results of this study: 

\begin{itemize}
    \item \gendirect\ was trained as described in \citetalias{Lai_2025_GenDIReCT} utilising the aperture coverage from datasets of 3C~279 and Centaurus~A. Once fully trained, validated, and applied on the real data, we find a high degree of consistency between the median \gendirect\ reconstruction and the appropriate reference reconstruction from the EHTC, as measured by the maximum normalised cross-correlation $\rho_{\rm{NX}} > 0.9$, where a score of unity denotes exact image correspondence. The data fidelity score of the best single reconstruction from \gendirect\ is $\chi^2_{\rm{CI}} \lesssim 2$.  
    \item We identify clusters of candidate image reconstructions from \gendirect\ by clustering image solutions based upon perceptual hashing algorithms. In both cases of 3C~279 and Centaurus~A, a significant portion of all reconstructions fall within a single cluster, while less populated clusters exhibit morphological diversity which are symptoms of the sparse aperture coverage VLBI image reconstruction, such as the phantom repetition of similar structures.
    \item By fitting the image reconstructions with parameterised image models, we construct quantitative measures of morphological characteristics which we compare between the EHTC reference images and the median \gendirect\ reconstruction. For 3C~279, we fitted three asymmetric Gaussian components to compare the intensity ratio between the core and ejecta ($I({\textrm{C0-1}})/I(\textrm{C1})$), the separation ($\abs{\textbf{C0} - \textbf{C1}}$), and angle ($\theta_{\textbf{C1} - \textbf{C0}}$) formed between the two components, as well as the position angle of the bright extended component (PA$_{\textbf{C0-1}}$). We find all measured features to be statistically similar between the EHTC ($I({\textrm{C0-1}})/I(\textrm{C1})=1.52\pm0.01$, $\abs{\textbf{C0} - \textbf{C1}} = 90.9\pm0.2\,\mu{\rm{as}}$, $\theta_{\textbf{C1} - \textbf{C0}} =-75.7 \pm0.1$ deg, PA$_{\textbf{C0-1}}=44.5\pm0.5$ deg) and \gendirect\ reconstructions ($I({\textrm{C0-1}})/I(\textrm{C1})=1.53\pm0.02$, $\abs{\textbf{C0} - \textbf{C1}} = 90.2\pm0.2\,\mu{\rm{as}}$, $\theta_{\textbf{C1} - \textbf{C0}} =-75.6 \pm0.1$ deg, PA$_{\textbf{C0-1}}=44.2\pm0.7$ deg), especially if taking into consideration a systematic position error at a fraction (5\%) of the clean beam dimensions. 
    \item For Centaurus~A, we fitted linear components to the two ridgelines and measured the intensity ratio ($R_{\rm s/n}$) as well as the jet opening angle ($\theta_{\rm s/n}$), defined by the angle formed between the two fitted lines. Both quantities were near-identical between the EHTC ($R_{\rm s/n}=1.5\pm0.2,\,\theta_{\rm s/n} = 12.2\pm0.1$ deg) and median \gendirect\ reconstructions ($R_{\rm s/n}=1.4\pm0.1,\,\theta_{\rm s/n} = 12.3\pm0.3$ deg). 
    \item We examine the relative motion of nuclear and ejecta components in 3C~279 across four days of EHT observations using \gendirect\ reconstructions on the combined dataset of the first two and latter two dates, quantifying the relative proper motion on sub-parsec scales. Despite performing our analysis on the image reconstructions rather than model-fitted visibilities, we find consistent estimates of the ejecta proper motion in both magnitude and direction, with a total angular displacement of $4.6\pm1.0\,\mu$as over $\simeq 5.39$ days, indicating apparent superluminal motion of $\beta_{\rm app} = 10\pm2$. This can be compared to the $4.4\pm0.9\,\mu$as displacement in the flux-weighted centroid of the three ejecta subcomponents in \citet{Kim_2020_3C279}.
\end{itemize}

The results of \gendirect\ on the datasets of 3C~279 and Centaurus~A show remarkable consistency with the EHTC image reconstructions. As an imaging algorithm conditioned on purely closure information, \gendirect\ offers a calibration-independent constraint on source morphology, which ultimately enhances the reliability of image reconstruction on sparsely sampled datasets. Moreover, we showcase the utility of novel machine learning algorithms in solving challenging inverse problems characteristic of radio interferometric imaging. This study presents, for the first time, independent image reconstructions from EHT datasets on 3C~279 and Centaurus~A as a demonstration of the \gendirect\ imaging method on real EHT datasets, paving the way for \gendirect\ to be applied on the event-horizon-scale imaging problem of M87 and Sgr A*. With potential to be expanded for dynamic and polarimetric imaging, the techniques underlying \gendirect\ have the potential to be more widely applied on imaging inverse problems throughout radio interferometry.

\section*{Acknowledgements}
We acknowledge Jae-young Kim and Michael Janssen for supplying the Event Horizon Telescope image reconstructions of 3C~279 and of Centaurus~A, respectively, which were for the comparative analysis. Inputs from Phil Edwards, Ron Ekers, and Mark Cheung are gratefully acknowledged. We acknowledge use of \textsc{GitHub Copilot}.

Software packages used in this study include \textsc{Numpy} \citep{Numpy_2011}, \textsc{Scipy} \citep{Scipy_2020}, \textsc{PyTorch} \citep{PyTorch}, \ehtim\ \citep{Chael_2018_ehtim}, \gendirect\ \citep{GenDIReCT_Zenodo}, \textsc{ClosureInvariants} \citep{CI_Package}, and \textsc{Matplotlib} \citep{Matplotlib_2007}.

\paragraph{Funding Statement}
None

\paragraph{Competing Interests}
None

\paragraph{Data Availability Statement}
The data underlying this article will be shared on reasonable request to the corresponding author. Code for \gendirect\ is publicly accessible on GitHub\footnote{\href{https://github.com/samlaihei/GenDIReCT}{https://github.com/samlaihei/GenDIReCT}}.

\paragraph{Ethical Standards}
The research meets all ethical guidelines, including adherence to the legal requirements of the study country.

\paragraph{Author Contributions}
Conceptualization: S.L; N.T. Methodology: S.L; F.D; N.T; I.W. Data curation: S.L. Data visualisation: S.L. Writing original draft: S.L.; N.T. All authors approved the final submitted draft.

\printendnotes

\bibliography{bibliography}
\appendix

\section{Perceptual Hashing} \label{appendix:phashing}
Perceptual hashing is a form of fingerprinting algorithm that is often applied on various types of multimedia, including images. Unlike cryptographic hashing techniques, perceptual hash algorithms are designed to minimise variation and preserve the resulting hash when the inputs are similar.

\begin{figure}
    \includegraphics[width=0.9\columnwidth]{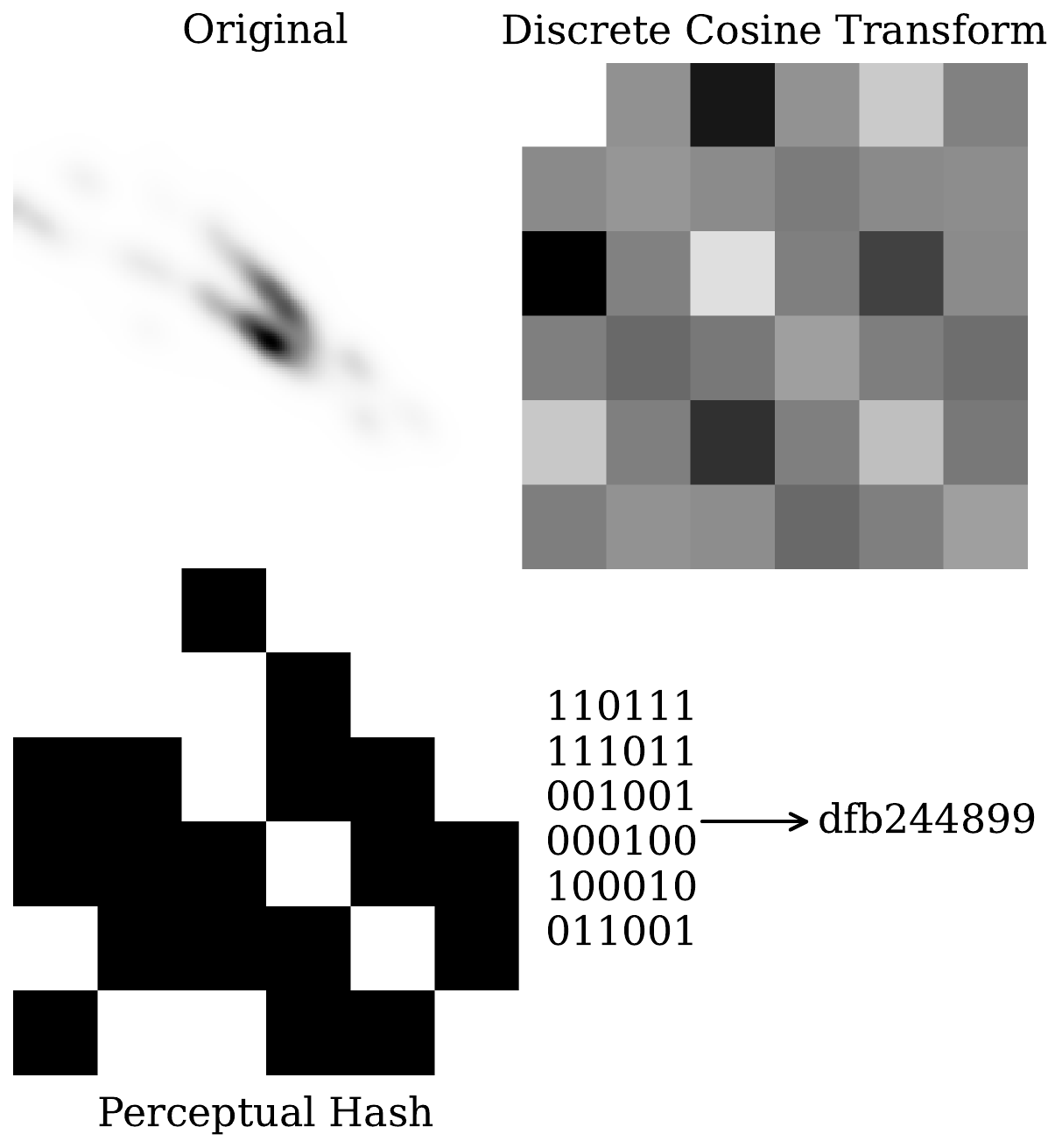}
    \caption[]{(Top left) Model image of Centaurus~A's jet \citep{Janssen_2021_CenA}. (Top right) Processed image after applying the discrete cosine transform on both axes and selecting the $6\times6$ box of the lowest frequency features. (Bottom left) The image is converted to binary by using the median value of the data as the threshold. White and black pixels correspond to 1 and 0, respectively. (Bottom right) Hexadecimal representation of the binary hash.}
    \label{fig:phash}
\end{figure}

To quantify image similarity across a generated sample for clustering purposes, we calculate the discrete cosine transform (DCT) hash for each reconstructed image. As illustrated in Figure \ref{fig:phash}, the DCT hash is a transformation which begins by computing the discrete cosine transform,
\begin{equation}
    X_k = \sum_{n=0}^{N-1} 2x_n\cos\left(\frac{\pi}{2N}(2n+1)k\right)\,,
\end{equation}
for each row and column. We then select the $s\times s$ box of pixels corresponding to the lowest frequency features, which are the least sensitive to noisy or small-scale variations. Finally, we convert the DCT image into binary using the median of the transformed data as the cutoff value. Optionally, we can identify the hash of any image by its hexadecimal representation as shown in the bottom right panel of Figure \ref{fig:phash}. 

Similarity is measured by the Hamming distance \citep{Hamming_1950}, defined as the number of bits for which the corresponding bit in the comparison sequence is flipped. In our implementation, we select $s=5$ and apply a Hamming distance threshold of $6$ for identifying members of a cluster unless stated otherwise. 

\section{Alternative reconstructions} \label{appendix:alternative-imaging}
Although \gendirect\ imaging does not depend on fine-tuned hyperparameters, such as selected priors or independently weighted regularisation terms, its result is not immune from pre-imaging or data processing considerations. In this appendix, we explore the outcome of different selections of the image reconstruction window and data processing strategies.

\begin{figure*}
	\includegraphics[width=0.8\textwidth]{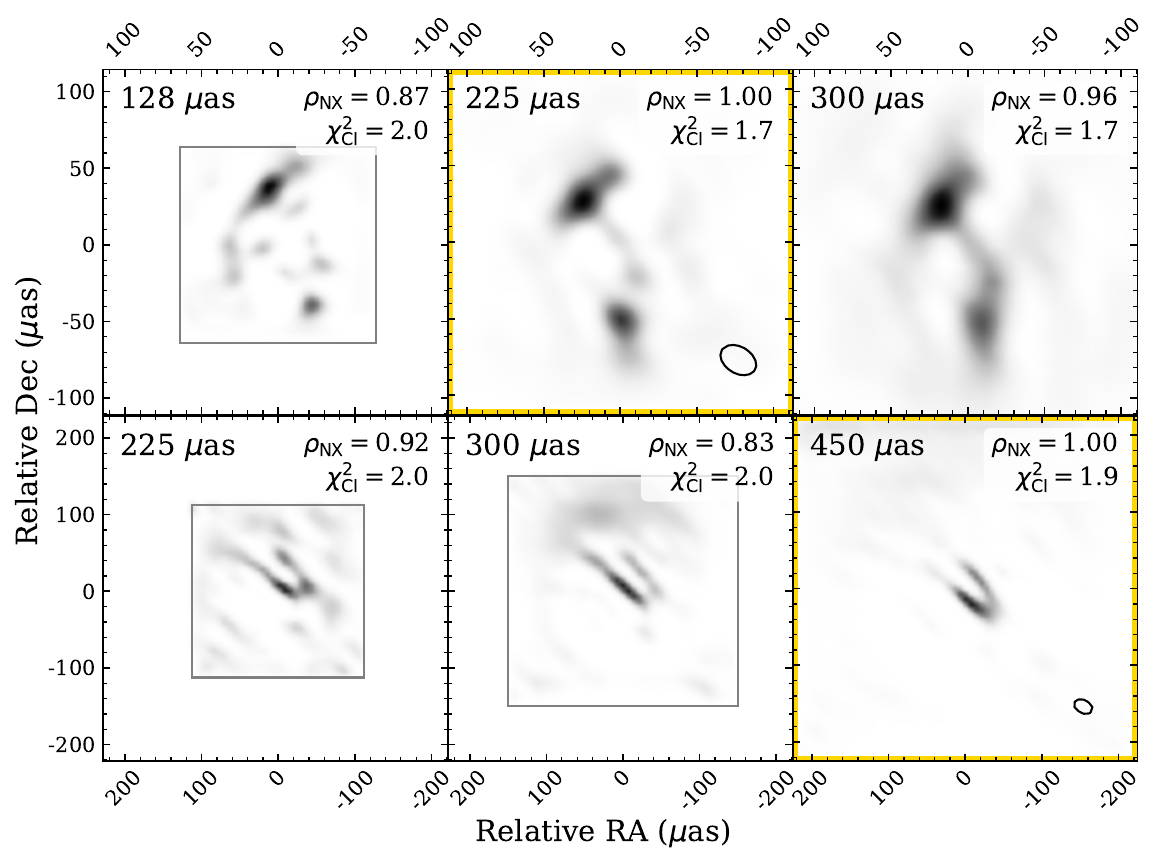}
    \caption[]{Median \gendirect\ reconstructions of 3C~279 (top) and Centaurus~A (bottom) at varying square fields-of-view from widths of (128, 225, 300 $\mu$as) for 3C~279 and (225, 300, 450 $\mu$as) for Centaurus~A. The fitted elliptical clean beam is plotted on the reconstruction that is used as reference for computing the relative $\rho_{\rm{NX}}$ and the reference image is also highlighted with a gold border. The best $\chi^2_{\rm{CI}}$ data fidelity score on individual \gendirect\ reconstructions is presented in each panel. For any reconstruction windows smaller than the reference field-of-view, we plot the edges of the reconstruction window.}
    \label{fig:fov-variation}
\end{figure*}

\subsection{Field-of-view effects} \label{appendix:fov-effects}
Due to the paucity of short and intermediate ($\lesssim 1\,{\rm{G}}\lambda$) baselines in the array, EHT observations exhibit limited sensitivity to extended structure on scales of $\gtrsim100\,\mu$as \citep[e.g.][]{Kim_2020_3C279}. As such, the EHT field-of-view is relatively compact. Consequently, we have carefully selected the image reconstruction field-of-view for 3C~279 and Centaurus~A based on the EHTC reference images to capture as much of the contributing emission sources while adhering to the spatial scales that fall within the sensitivity range of the EHT. However, as we have described regarding the validation tests in Section \ref{sec:validation-performance}, in cases where significant emission lies outside the reconstruction field-of-view, limiting the image reconstruction window can produce biased and unfaithful reconstructions. Therefore, in this section, we train \gendirect\ for the 3C~279 and Centaurus~A datasets with modified reconstruction windows to ensure consistency in the image reconstruction output.

In Figure \ref{fig:fov-variation}, we present the median reconstructions from \gendirect\ on 3C~279 with the 11 April 2017 dataset and Centaurus~A, each with three different reconstruction windows. For 3C~279, we have chosen $128\times128\,\mu{\rm as}^2$, $225\times225\,\mu{\rm as}^2$, and $300\times300\,\mu {\rm as}^2$. On Centaurus~A, the original $450\times450\,\mu{\rm{as}}^2$ field-of-view is already large relative to the EHT array's sensitivity. Therefore, we selected smaller alternatives of $225\times225\,\mu{\rm as}^2$ and $300\times300\,\mu{\rm as}^2$ to visualise the edge-brightened jet sheath in closer detail. We present the relative $\rho_{\rm{NX}}$ image correspondence score between the median \gendirect\ reconstruction in each of the alternative fields-of-view and the default median image used in this study ($225\,\mu$as for 3C~279 and $450\,\mu$as for Centaurus~A), which is also highlighted with the gold border. 

As described in Section \ref{sec:results}, the image reconstruction artifacts northwards of the edge-brightened Centaurus~A jet are clipped to focus the $\rho_{\rm{NX}}$ comparison on the central feature. For any fields-of-view smaller than the reference, we illustrate the edges of the reconstruction windows for visual clarity as all reconstructions are presented in the reference field-of-view. Additionally, we compute the best $\chi^2_{\rm{CI}}$ of the individual \gendirect\ reconstructions in each field-of-view, showing that the widths of $225\,\mu$as for 3C~279 and $450\,\mu$as for Centaurus~A result in the optimum $\chi^2_{\rm{CI}}$ compared to distinct field-of-view alternatives. All of the panels showcasing alternative reconstruction windows in Figure \ref{fig:fov-variation} are closely consistent ($\rho_{\rm{NX}} > 0.80$) with the reference image reconstruction without significant deviations in the overall reconstructed image morphology, lending enhanced credibility to the imaging results obtained in this study. 

\begin{figure}
	\includegraphics[width=0.95\columnwidth]{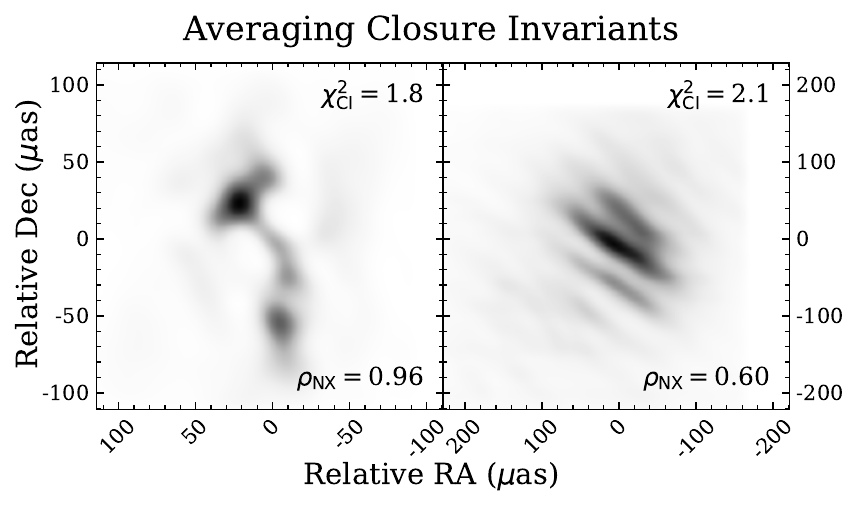}
    \caption[]{Median \gendirect\ reconstructions of 3C~279 (left) and Centaurus~A (right) from the closure invariant averaged dataset. The best single-reconstruction $\chi^2_{\rm{CI}}$ is shown on the top-right of each panel and the $\rho_{\rm{NX}}$ compared to the median visibility-averaged \gendirect\ reconstruction is displayed in the bottom-right corner of each panel.}
    \label{fig:CI-avg-results}
\end{figure}

\subsection{Alternative data aggregation strategy} \label{appendix:ci-avg}

As described in Section \ref{sec:imaging}, the imaging results in this section are produced by averaging visibility information across $\sim5$-minute scans to compress the dataset into a more manageable volume and improve SNR. However, while station-based multiplicative corruptions are removed by the closure invariants construction, visibilities can be further corrupted by time-dependent errors across the duration of the scan, the effect of which would propagate into the closure invariants. To mitigate the time-dependent effects, it's possible to compute the closure invariants in the $10\rm{s}$-binned cadence of the public dataset. In \citetalias{Lai_2025_GenDIReCT}, it was demonstrated empirically that pre-computing and averaging the closure invariants at a higher temporal resolution is generally a more robust method of data aggregation than averaging the visibilities on the same timescale prior to computing closure invariants in the presence of noise. However, some exceptions were identified, such as when the total flux density is low, then under standard SEFDs, there is little difference between both strategies. Additionally, when the total flux density is on the order of $\sim1$ Jy and the visibility phase is well-calibrated, then averaging visibilities is superior at preserving the true closure invariant quantities than pre-computing closure invariants at a higher temporal resolution prior to averaging. Indeed, if the visibilities are well-calibrated, averaging the visibilities produces a more robust result for any source with $\sim1$ Jy or brighter. On the other hand, if the calibration is uncertain, averaging the closure invariants is superior and its relative advantage grows further with source brightness. 

For Figure \ref{fig:CI-avg-results}, we train new \gendirect\ models with the closure invariant averaging strategy and present the resulting median image reconstructions of 3C~279 on the 11 April 2017 dataset and Centaurus~A, alongside the best single-reconstruction $\chi^2_{\rm{CI}}$ data metric and $\rho_{\rm{NX}}$ of the median closure invariant averaging reconstruction relative to the corresponding visibility averaging dataset image presented in Section \ref{sec:results}. We find that for 3C~279, the resulting median reconstruction from the closure invariant averaged dataset is $\rho_{\rm{NX}}=0.96$ relative to the visibility averaged dataset with statistically similar performance on $\chi^2_{\rm{CI}}$. However, the quality of the image reconstruction of Centaurus~A is degraded. Although the general tuning fork morphology is visibly identifiable, the median reconstruction shows parallel features similar to structures in the dirty image, which decreases the image correspondence with the visibility averaged reconstruction to $\rho_{\rm{NX}} = 0.6$. The total compact flux density of Centaurus~A is $\sim2$ Jy \citep{Janssen_2021_CenA} and after averaging the closure invariants, the median closure invariant SNR is $\sim 2$, which is likely too low for the \gendirect\ reconstruction to be considered reliable, even on tests with synthetic data \citepalias{Lai_2025_GenDIReCT}. For comparison, the median closure invariant SNR from the visibility-averaged dataset is $\sim5$. Thus, we find that due to its low compact flux density, the phase-calibrated dataset of Centaurus~A falls within the range where visibility averaging results in a superior reconstruction. Comparably, we find minimal difference in reconstruction quality of 3C~279 due to its high SNR. We further anticipate that with a compact flux density of $\sim1$ Jy, the calibrated datasets of M87 are expected to display similar characteristics as Centaurus A in this respect.

\begin{figure}
	\includegraphics[width=0.9\columnwidth]{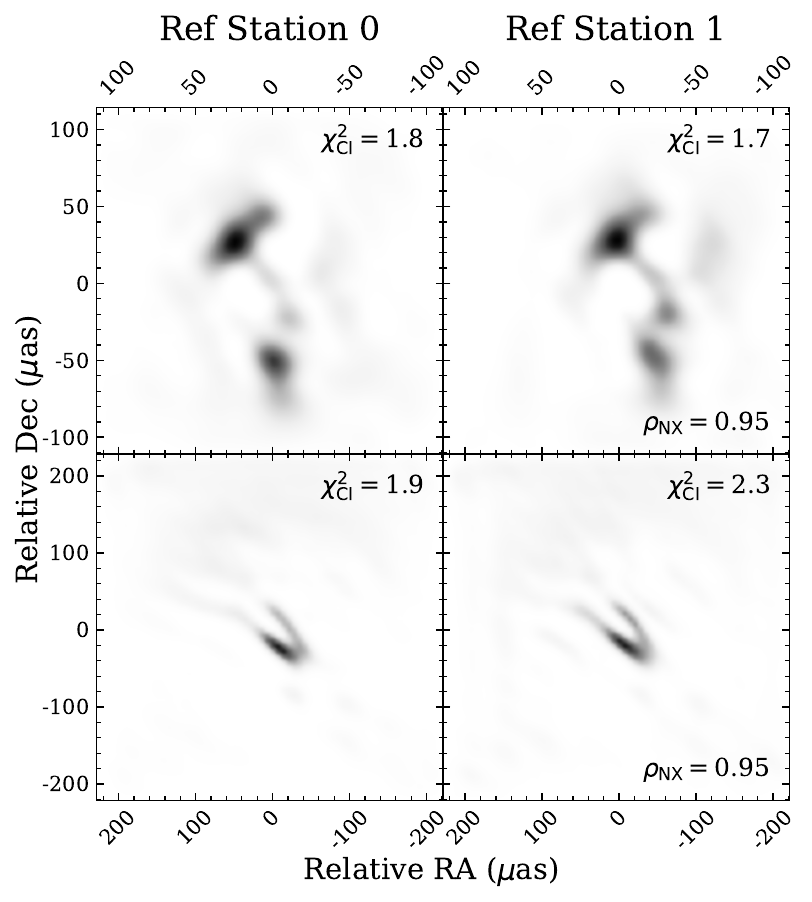}
    \caption[]{Comparison illustration of median \gendirect\ reconstructions of 3C~279 (top) and Centaurus~A (bottom) with different choices of reference station in each scan ($0-$ most sensitive station, $1-$ second-most sensitive station). Both reconstructions of each target are structurally consistent to $\rho_{\rm{NX}}=0.95$.}
    \label{fig:baseid-results}
\end{figure}

\subsection{Alternative closure triad reference stations} \label{appendix:baseid}

One can identify all independent triangles for an interferometer by fixing a base vertex or reference station and selecting all triangles containing the vertex \citep{TMS,Thyagarajan_2022_CI}. A different selection of the reference station produces a different set of closure invariants carrying equivalent information. In this section, we investigate the choice of the reference station on the resulting image reconstruction by alternatively selecting the second most sensitive station in each scan instead of the most sensitive station. With this selection, the overall SNR of the closure invariants decreases due to the less sensitive base station and increased covariances. 

We illustrate the median \gendirect\ reconstruction for both 3C~279 (top) and Centaurus~A (bottom) in Figure \ref{fig:baseid-results} alongside their respective $\chi^2_{\rm{CI}}$ data fidelity scores and relative $\rho_{\rm{NX}}$. The left column reconstructions utilise the most sensitive station as the reference in each scan, while the right column reconstructions utilise the second most sensitive station. For both targets, we find a high degree of image consistency with either reference station selection, with $\rho_{\rm{NX}}=0.95$ correspondence between the median \gendirect\ reconstructions. As for the data fidelity, we note that with reference station 1, the reconstruction performance on $\chi^2_{\rm{CI}}$ is degraded for Centaurus~A, but not for 3C~279. The reason is likely that when the less sensitive station is used consistently in construction of closure triads, the overall SNR of the resulting closure invariants also decreases, which affects the observations of Centaurus~A, observed with a median visibility $\rm{SNR}\sim5$, disproportionately compared to observations of 3C~279, where the median visibility is significantly greater ($\rm{SNR}\sim 23$).

\end{document}